\documentclass[12pt,a4paper]{article}
\pdfoutput=1
\usepackage{jheppub,tcolorbox}
\usepackage[utf8]{inputenc}
\usepackage{amsmath,amssymb}
\usepackage{float}
\usepackage{amsfonts}
\usepackage{dsfont}
\usepackage{color}
\usepackage{varwidth}
\usepackage[framemethod=tikz]{mdframed}
\usepackage{epigraph}
\usepackage{tikz}

\newcommand{\Arrow}[1]{%
\parbox{#1}{\tikz{\draw[<->](0,0)--(#1,0);}}
}

\usepackage{tikz} 
\usetikzlibrary{decorations.markings} 
\usetikzlibrary{decorations.text} 
\usetikzlibrary{positioning} 
\usetikzlibrary{backgrounds} 
\usetikzlibrary{fit} 
\usetikzlibrary{calc} 
\usetikzlibrary{shapes} 

\tikzset{dot/.style={draw,circle,inner sep=.7pt,fill,node
    distance=1cm}} 
\tikzset{dot1/.style={draw,circle,inner sep=.7pt,fill}} 
\tikzset{triangle/.style={draw,regular polygon, regular polygon
    sides=3}} 
\tikzset{->-/.style={decoration={
  markings,
  mark=at position .5 with {\arrow{>}}},postaction={decorate}}} 
\tikzset{-<-/.style={decoration={ 
  markings,
  mark=at position .5 with {\arrow{<}}},postaction={decorate}}}

\newcommand\be{\begin{equation}}
\newcommand\ee{\end{equation}}
\newcommand\bea{\begin{eqnarray}}
\newcommand\eea{\end{eqnarray}}

\begin{document}

\begin{titlepage}
\renewcommand{\thefootnote}{\fnsymbol{footnote}}


\vspace*{2.0cm}

\begin{center}
{\textbf{\huge Spontaneous Symmetry Breaking in Tensor Theories 
}}%
\end{center}
\vspace{1.0cm}

\centerline{
\textsc{\large{P. Diaz}} \footnote{pablodiazbe@gmail.com},
\textsc{\large{J. A.  Rosabal}} \footnote{j.alejandro.rosabal@gmail.com}
}

\vspace{0.6cm}

\begin{center}
\it Fields, Gravity \& Strings @ {\rm CTPU}, Institute for Basic Science, \\
55, Expo-ro, Yuseong-gu, Daejeon, Korea, 34126.
\end{center}

\vspace*{1cm}

\centerline{\bf Abstract}

\begin{centerline}
\noindent
In this work we  study  spontaneous symmetry breaking patterns in tensor models. We focus on the patterns which lead to effective matrix theories transforming in the adjoint of $U(N)$. We find the explicit form of the Goldstone bosons which are organized as matrix multiplets in the effective theory. The choice of these symmetry breaking patterns is motivated by the fact  that, in some contexts, matrix theories are dual to gravity theories. Based on this, we aim to build a bridge between tensor theories, quantum gravity and holography.

\end{centerline}
\thispagestyle{empty}
\end{titlepage}

\setcounter{footnote}{0}

\tableofcontents
\newpage


\section{\label{sec:1}Introduction}

Nowadays, there is a wave of  excitement in the tensor field community \cite{Gurau:2009tw, Gurau:2011xp, Gurau:2010ba, Gurau:2011aq, Gurau:2011xq,  Klebanov:2017nlk, Giombi:2017dtl, Bulycheva:2017ilt, Giombi:2018qgp, Klebanov:2018fzb, Ferrari:2017jgw,Geloun:2013kta, BenGeloun:2017vwn,Itoyama:2017emp,Mironov:2017aqv,Itoyama:2017wjb,Itoyama:2018but,Diaz:2017kub,Diaz:2018xzt,Diaz:2018zbg} 
triggered by the remarkable solvability of certain  models in the large $N$ \cite{Gurau:2009tw, Gurau:2011xp, Gurau:2010ba, Gurau:2011aq, Gurau:2011xq,  Klebanov:2017nlk, Giombi:2017dtl, Bulycheva:2017ilt, Giombi:2018qgp,Klebanov:2018fzb, Ferrari:2017jgw}. Especially since they were discovered to be related to holography\cite{Witten:2016iux}, in the context of SYK duality \cite{Azeyanagi:2017drg,Yoon:2017gut,Yoon:2017nig,Narayan:2017hvh,Nosaka:2018iat,Jevicki:2016bwu,Jevicki:2016ito,Das:2017pif,Das:2017wae}. 
These models involve very particular ``melonic'' interactions, which permit to exactly solve the Schwinger-Dyson equation. It is clear that those models do not exhaust the vast spectrum of tensor theories. Restricting ourselves to those solvable models, many physically relevant cases could be left out. 

A more general scenario will presumably involve interaction terms that would take the theory out of its large $N$  solvability.   
In order to tackle generic tensor theories other techniques will be necessary.  It is likely that the addition of interaction terms could provide theories with spontaneous symmetry breaking (SSB). 

The purpose of this paper is to kinematically explore the vacuum in tensor theories. For this, we study SSB scenarios, especially when they lead to effective (multi-)matrix theories transforming in the adjoint of $U(N)$, which are the Goldstone bosons.  The choice of such SSB patterns is mainly motivated by the fact that matrix theories are  related to quantum gravity and holography.  

Instead of aiming to fully solve the tensor theory, in this work we shall initiate the programme of studying the effective (matrix)  theory arising from SSB. Assuming that at low energy the Goldstone bosons associated with SSB are the only massless degrees of freedom, the dynamics of the tensor theory could be handled in terms of matrix theories. Actually, constructing phenomenological actions for Goldstone bosons has been very successful, for instance the chiral symmetry breaking in QCD. Conversely, it could be the case that computations in a given matrix theory are hard but the uplifted tensor theory is easier to handle.

Matrix fields have been proven to furnish a  dual description \cite{Banks:1996vh, DiFrancesco:1993cyw,Berenstein:2004kk, deMelloKoch:2007rqf} of a quantum gravity theory. In this work, we see how matrix fields come up as effective degrees of freedom of a tensor theory. It is reasonable to think that the tensor theory could actually be describing the same dual theory, although at a higher energy regime. Schematically,
\begin{equation*}
\begin{array}{ccc}
\boxed{\text{Tensor Theory}}&\Arrow{2cm} &\boxed{\text{Dual theory 
(higher energy)?}}\\
\text{SSB}\Bigg\downarrow &&\Bigg\downarrow ?\\
\boxed{\text{Matrix Theory}}&\Arrow{2cm} &\boxed{\text{~~~~~~~~~~~Dual theory~~~~~~~~~~~~}}
\end{array}
\end{equation*}
The main result of the paper is \eqref{mainresult} with \eqref{2summands}. It encompasses the identification of the matrix fields as multiplets of Goldstone bosons which come from SSB, together with the choice of the parameter which induces and specifies the SSB pattern. Explicitly,
\begin{equation*}
(B^{\alpha,k'})^i_j(x)=\frac{1}{2}\Big(\overline{v}^{i_1\dots i_d}_{j_1\dots j\dots j_d}\Phi^{j_1 \dots i \dots j_d}_{i_1\dots i_d}(x)-v^{j_1\dots i\dots j_d}_{i_1\dots i_d}\overline{\Phi}^{i_1\dots  i_d}_{j_1\dots j \dots j_d}(x)\Big),
\end{equation*}
with
\begin{equation*}
v^{i_1\dots i_d}_{j_1\dots j_d}=v_\sigma\delta^{i_1\dots i_d}_{j_{\sigma(1)}\dots j_{\sigma(d)}}+v_{\sigma'}\delta^{i_1\dots i_d}_{j_{\sigma'(1)}\dots j_{\sigma'(d)}}, 
\end{equation*}
where $B^{\alpha,k'}(x)$ are $N\times N$-matrix fields transforming in the adjoint of $U(N)$. There will be several matrix fields labeled by $(\alpha,k')$, the meaning of which will be explained later. The tensor parameter which induces SSB is $v^{i_1\dots i_d}_{j_1\dots j_d}$, and $(\sigma,\sigma')$ are the only two permutations which are sufficient to specify the pattern. The tensor field is denoted by $\Phi^{i_1\dots i_d}_{j_1\dots j_d}(x)$. 

As usual in color tensor theories \cite{Gurau:2009tw, Gurau:2011xp, Gurau:2010ba, Gurau:2011aq, Gurau:2011xq}, there is a different $U(N)$ symmetry group associated to every index. The full symmetry group will be defined as $G_{d\bar{d}}(N)=\prod_{k=1}^d [U_k(N)\times U_{\bar{k}}(N)]$.  
In this work we will not consider any definite form of the interaction potential. Instead, we will assume that there exists theories that present SSB patterns into diagonal subgroups of $G_{d\bar{d}}(N)$. 

In the classical example of  chiral symmetry breaking,  the scalar field is $\Phi^i_j(x)$, the pattern is $U(N)_L\times U_R(N)$ into the diagonal subgroup $U(N)$, and the $v$ tensor is $v^i_j=v\delta^i_j$  \cite{Coleman:1980mx}. Unlike chiral symmetry breaking, the SSB patterns for tensor models present a much richer structure. Because of the tensor nature of the field $\Phi^{i_1\dots i_d}_{j_1\dots j_d}(x)$, the generalization of the chiral symmetry breaking will correspond to the different diagonal subgroups that can be made by suitably splitting $G_{d\bar{d}}(N)$. Accordingly, each SSB pattern is associated with a partition of $d$.

The SSB patterns into the diagonal subgroups are induced by the invariant tensor $v$. If we wish to break into diagonal subgroups, the tensor $v$ should be a linear combination of a product of $d$ Kronecker deltas.  There are $d!$ different delta monomials (driven by all possible permutations of $d$ indices) and a linear combination $v$ would, in principle, involve all of them. This would be the natural generalization of the choice $v^i_j=v\delta^i_j$ for chiral symmetry breaking.
Remarkably enough, the suitable choice of {\it only two} monomials, as in equation \eqref{2summands}, can induce {\it whatever} SSB pattern of the type we are interested in, and it holds for tensor theories of {\it any} rank.

The above statement is rigorously proven in section \ref{MS}. For that, we first note that a single delta monomial in the $\epsilon$-term halves $G_{d\bar{d}}(N)$ into diagonal subgroups. The linear combination of two (or more) different monomials reduces the remaining symmetry group even more, to the intersection among the subgroups associated with each monomial. These observations, inspired us to build diagrams that represent  the intersection of the groups that result from the combination of two delta monomials. Conversely, from a given diagram one can read off the two constituent monomials, and its SSB pattern. It turns out that {\it any SSB pattern can be realized in the diagrams}. Therefore, any SSB can be induced by only two monomials \eqref{2summands}.

In order to explore the SSB patterns and, more importantly, to identify the Goldstone bosons as in \eqref{mainresult}, we introduce the $\epsilon$-term technique in the path integral  for tensor theories. As in the classical examples \cite{Matsumoto:1974nt, Matsumoto:1974gp}, the role of the $\epsilon$-term in tensor theories is to restrict the symmetry of the vacuum. Besides, it allows us to derive the Ward-Takahashi (WT) identities, which relate different Green functions of the theory with SSB. The WT identities reveal the explicit form of the Goldstone bosons in terms of the $2d$-rank tensor field $\Phi(x)$.  \\

There is an increasing interest in clarifying the role of tensor theories in the context of holography \cite{Witten:2016iux,deMelloKoch:2017bvv}, see also \cite{Klebanov:2018fzb} and references therein. We hope that the study of SSB in tensor models helps in this direction. Moreover, in more general terms, SSB could be also implemented  in theories which are known to be  solvable in the large $N$. The theories do no need to be perturbatively stable, they just require the existence of stationary points of the potential, see, for instance, section $19.3$ of \cite{Weinberg:1996kr} and p. 246 of  \cite{Polchinski}.  There is no need, either, for the symmetry group to be unitary. Actually, since SYK involves real fields, $O(N)$ theories are becoming popular. As a remark, there is a resemblance between the zero  temperature limit of certain $O(N)$ theories \cite{Choudhury:2017tax}  and the material presented in this paper.       \\

The paper is organized as follows. As a warm-up, in section \ref{scalar0} we introduce the $\epsilon$-technique by reviewing the complex scalar case and its kinematical SSB \cite{Matsumoto:1974nt}. We also fix the notation and conventions we will use throughout the paper. In section \ref{SSpatterns} we explore the SSB patterns which lead to diagonal subgroups of $G_{d\bar{d}}(N)$; we introduce the ``intersection-diagrams'' which correspond to the different SSB patterns; and in section \ref{MS}, we use the diagrammatic correspondence to prove the statement that leads to equation \eqref{2summands}. In section \ref{general formalism}, we develop the general formalism of SSB in tensor theories. We implement the $\epsilon$-term technique in these theories as a generalization of the method applied for the scalar field, reviewed in section \ref{scalar0}. As far as we know, this has not been reported in the literature yet. Using this formalism (in particular, the WT identities), we identify the Goldstone bosons associated to a particular SSB pattern and we present our main result in equation \eqref{mainresult}. The classical example of chiral symmetry breaking is treated in our formalism in section \ref{E2}. In order to illustrate the general treatment in a less trivial case, we present the 4-rank tensor example in section \ref{E4}. After the conclusions, in appendix \ref{tetraap}, we show an example of how this formalism applies to other symmetry groups and different SSB patterns. In particular, by minimizing the potential, we find the SSB patterns and derive the Goldstone bosons in the model with tetrahedral interaction \cite{Klebanov:2016xxf}.

\section{General setup}\label{scalar0}
As a warm-up  we will first review SSB in the case of a complex scalar field \cite{Matsumoto:1974nt}. We shall show the  $\epsilon$-term technique in the path integral to induce
SSB. It allows us to identify the Goldstone boson by means of the WT identities. The technique has a straightforward extension to higher rank tensor theories.  
\subsection{Complex scalar field and its SSB}

For the study of the SSB in the complex scalar field theory we will require an action  invariant under the phase transformation 
\be
\Phi(x)\rightarrow\text{e}^{\text{i}\alpha}\Phi(x), \label{sym1}
\ee
being $\alpha$ an arbitrary real constant, i.e., 
\be
S\big[\Phi(x)\big]=\int d^dx {\cal L}\big[\text{e}^{\text{i}\alpha}\Phi(x)\big]=\int d^dx {\cal L}\big[\Phi(x)\big],
\ee
with ${\cal L}$ the Lagrangian.
In order to derive the WT identities for the SSB we define the generating functional 
\be
Z[\overline{J},J]=\frac{1}{\cal N}\int D\Phi D\overline{\Phi}\text{exp}\Big[\text{i}S\big[\Phi(x)\big]+\text{i}\int d^dx\big(\overline{J}(x)\Phi(x)+J(x)\overline{\Phi}(x)\big) \Big], \label{genefunc}
\ee
where the functional  integration is restricted to a vanishing field  at infinity. We suppose that the measure is also invariant under (\ref{sym1}), and ${\cal N}=Z[0,0]$. In contrast, we could define the generating functional with different boundaries conditions, i.e., with a constant field at infinity. This change in the usual boundary conditions  leads to the SSB.  In order to implement the  boundary conditions in the path integral we can define the new generating functional $Z_{\epsilon}[\overline{J},J]$,
\begin{multline}
Z_{\epsilon}[\overline{J},J]=\frac{1}{\cal N}\int D\Phi D\overline{\Phi}\text{exp}\Big[\text{i}S\big[\Phi(x)\big]+\text{i}\int d^dx\big(\overline{J}(x)\Phi(x)+J(x)\overline{\Phi}(x)\big)\\
-\epsilon\int d^dx|\Phi(x)-v|^2 \Big], \label{Wepsilon}
\end{multline}
where the limit $\epsilon\rightarrow 0$ is understood at the end of a given calculation and  ${\cal N}=Z_{\epsilon}[0,0]$.

To elucidate the role of the $\epsilon$-term  let us specialize (\ref{Wepsilon}) on the  Lagrangian ${\cal L}=\frac{1}{2}\partial_{\mu}\Phi\partial^{\mu}\overline{\Phi}-V\big[\Phi,\overline{\Phi}\big]$. The generating functional in this case can be computed as
\be
Z_{\epsilon}[\overline{J},J]=\text{exp}\Big[-\text{i}\int d^dxV\big[ -\text{i}\frac{\delta}{\delta J(x)}, -\text{i}\frac{\delta}{\delta \overline{J}(x)} \big]\Big] Z_{0,\epsilon}[\overline{J},J], \label{genefuc1}
\ee
where $Z_{0,\epsilon}[\overline{J},J]$ is (\ref{Wepsilon}) with the free Lagrangian.

Now we can solve the integral using the  condition of stationary phase for the free action. Proceeding in the standard fashion we define $\Phi(x)=\Phi_{cl}(x)+\chi(x)$, where 
\be
\Box \Phi_{cl}(x)=\text{i}\epsilon\big(\Phi_{cl}(x)-v\big)+J(x),
\ee
we find
\be
\Phi_{cl}(x)=v+\int d^dy(\Box-\text{i}\epsilon)^{-1}_{xy}J(y).
\ee
Now (\ref{Wepsilon}) takes the form
\be
Z_{0,\epsilon}[\overline{J},J]=\text{exp}\Big[\text{i}\int d^dx\big(v \overline{J}(x)+\overline{v}J(x)\big)\Big]\text{exp}\Big[\int d^dx\int d^dy J(x)(\Box-\text{i}\epsilon)^{-1}_{xy}\overline{J}(y) \Big]. \label{Wepsilon1}
\ee
Notice that   $\langle \Phi(x) \rangle_{\epsilon}= v$ and $\langle \overline{\Phi}(x) \rangle_{\epsilon} = \overline{v}$ for the free theory,  while for the interacting one the only claim we can make so far is $\langle \Phi(x) \rangle_{\epsilon}\neq 0$ and $\langle \overline{\Phi}(x) \rangle_{\epsilon} \neq 0$. The subindex $\epsilon$ in the expectation values indicates that  they are taken with respect to $Z_{\epsilon}[\overline{J},J]$.  

For obtaining the  WT identities we make use of the invariance of the generating function (\ref{Wepsilon}), under the transformation (\ref{sym1}). After implementing (\ref{sym1}) in the generating functional,  we compute 
\be
\frac{\partial}{\partial \alpha}Z_{\epsilon}[\overline{J},J]=0, \label{deri}
\ee
and we obtain
\be
\langle \Phi(x)\rangle_{\epsilon}=-\epsilon \int d^dy\langle \Phi(x), \overline{v}\Phi(y)-v\overline{\Phi}(y)\rangle_{\epsilon}. \label{WI}
\ee
We will  identify the Goldstone bosons by doing certain linear combinations  of derivatives of the form (\ref{deri}). This way the expressions  are easily generalizable to the non abelian and higher rank tensors cases. Defining the combinations 
\be
\Big[\overline{v}\frac{\partial}{\partial \alpha}Z_{\epsilon}[\overline{J},J]\pm v\frac{\partial}{\partial \alpha}\overline{Z}_{\epsilon}[\overline{J},J]\Big]=0 \label{deri2},
\ee
we get 
\be
\langle \overline{v} \Phi(x)+v \overline{\Phi}(x)\rangle_{\epsilon}=-\epsilon \int d^dy\langle \overline{v} \Phi(x)-v \overline{\Phi}(x), \overline{v}\Phi(y)-v\overline{\Phi}(y)\rangle_{\epsilon}, \label{WI1}
\ee
and
\be
\langle \overline{v} \Phi(x)-v \overline{\Phi}(x)\rangle_{\epsilon}=-\epsilon \int d^dy\langle \overline{v} \Phi(x)+v \overline{\Phi}(x), \overline{v}\Phi(y)-v\overline{\Phi}(y)\rangle_{\epsilon}. \label{WI2}
\ee
Defining 
\bea
\varphi(x) & = & \overline{v} \Phi(x)+v \overline{\Phi}(x),\\
B(x) & = & \overline{v} \Phi(x)-v \overline{\Phi}(x),
\eea
we can rewrite (\ref{WI1}) and (\ref{WI2}) in a more suggestive form
\be
\langle \varphi(x)\rangle_{\epsilon}=-\epsilon \int d^dy\langle B(x) , B(y) \rangle_{\epsilon}, \label{WI11}
\ee
and
\be
\langle B(x) \rangle_{\epsilon}=-\epsilon \int d^dy\langle \varphi(x), B(y) \rangle_{\epsilon}. \label{WI21}
\ee

The full propagator, considering the interactions,  in (\ref{WI11}) is given by 
\be\label{propagatorB}
\langle B(x) , B(y) \rangle_{\epsilon}=\text{i}\int\frac{d^d p}{(2\pi)^d}\frac{Z_B}{p^2-m_{B}^2+\text{i}\epsilon a_B}\text{e}^{\text{i}p\cdot(x-y)}+(\text{regular contributions}),
\ee
where we have assumed  at first that  the $B(x)$ field could be massive, and $Z_B$ and $a_B$ are renormalization constants.
Now, we plug the propagator \eqref{propagatorB} in  (\ref{WI11}). After integration, (\ref{WI11}) reduces to 
\be
\langle \varphi(x)\rangle_{\epsilon}=\frac{\text{i}\epsilon Z_B}{ -m^2_B+\text{i}\epsilon a _B}. \label{WI111}
\ee
Now, since we assume that the one point function does not vanish, we immediately conclude that $m_B=0$. The field $B(x)$ is the Goldstone boson associated to the broken symmetry (\ref{sym1}). This way we have illustrated how, when SSB occurs, the Goldstone boson is identified using the $\epsilon$-term technique.

\subsection{Conventions and notation}
The theories we are considering in this paper are built on bosonic $2d$-rank tensors, $ \Phi_{j_1\dots j_d}^{k_1\dots k_d}(x)$, transforming under the symmetry group
\be \label{gdd}
G_{d\bar{d}}(N)=\prod_{k=1}^d [U_k(N)\times U_{\bar{k}}(N)],
\ee
where $U_k(N)$ and  $U_{\bar{k}}(N)$ are different groups, acting on the upper and lower indices, respectively. A general element of 
$G_{d\bar{d}}(N)$ is
\be\label{genericgdd}
(g_1,\dots, g_d,g_{\bar{1}},\dots, g_{\bar{d}}).
\ee

A field with one index downstairs will transform under $U(N)$. For convenience we define the conjugate field with the index upstairs, which will transform under the same $U(N)$. The transformation law will be 
\be\label{trans1}
\Phi'_i=g_i^l\Phi_l\rightarrow \Phi g ,\quad g\in U(N),
\ee 
\be\label{trans2}
\overline{\Phi}'^i=g_l^{\dagger i}\overline{\Phi}^l\rightarrow g^\dagger \overline{\Phi},\quad g\in U(N).
\ee 
Note that, with this convention, the field conjugate $\overline{\Phi}$  will transform with the transpose conjugate\footnote{The transpose conjugate of $g^i_j$ is $(g^\dagger)^j_i$. It is worth checking up the consistency of \eqref{trans1} and \eqref{trans2}, 
\be
\overline{ \Phi g} \nonumber \rightarrow \overline{g_i^l\Phi_l} 
=\overline{g}_i^l\overline{\Phi}^l =(g^{\dagger})_l^i\overline{\Phi}^l  \rightarrow g^{\dagger}\overline{\Phi}.
\ee}.

This way, the contraction $\Phi\cdot \overline{\Phi}=\Phi_i\overline{\Phi}^i$ is invariant under $U(N)$. Note that if we had associated the field and the field conjugate
both with the index downstairs the transformation law for the conjugate field would have been $\overline{g}$ instead of $g^\dagger$.

If the index of the field is upstairs, then the field and conjugate field transform as
\begin{eqnarray}
\Phi'^i&=&g^{\dagger i}_l\Phi^l \rightarrow g^{\dagger} \Phi, \nonumber \\
\overline{\Phi}'_i&=&g_i^{l}\overline{\Phi}_l \rightarrow \overline{\Phi} g,\quad g\in U(N),
\end{eqnarray}
so, $\overline{\Phi}_i\Phi^i$ is also an invariant. With this notation
\begin{itemize}
\item Upstairs indices always transform with the adjoint matrix. 
\item Contractions are always between downstairs and upstairs indices. The matrix multiplication is shown in the transformation laws above.
\end{itemize}

For a field with two indices we will have
\begin{eqnarray}\label{2ind}
\Phi'^i_j&=&(g_1^\dagger)^{i}_l(g_{\bar{1}})^k_j\Phi^l_k,\nonumber \\
\overline{\Phi}'^i_j&=&(g_{\bar{1}}^\dagger)^{i}_l(g_1)^k_j\overline{\Phi}^l_k,\quad g_1,g_{\bar{1}}\in U_1(N),  U_{\bar{1}}(N).
\end{eqnarray}
Notice that $g_1$ and $g_{\bar{1}}$ are two different elements, which belong to two different groups. We could have called them $g_1$ and $g_2$, for instance, but we find the notation ``1'' and ``$\bar{1}$'' convenient for the cases where more indices are involved. With this transformation law the contraction $\Phi\cdot\overline{\Phi}=\Phi^i_j\overline{\Phi}^j_i$ is invariant under $U_1(N)\times \overline{U}_{\bar{1}}(N)$. 

The fields that we will consider are, in general, higher rank tensors  $\Phi_{j_1\dots j_d}^{i_1\dots i_d}(x)$, which transform under the group $G_{d\bar{d}}(N)$, defined in \eqref{gdd}. They have $d$ indices upstairs and $d$ indices downstairs, each index transforming with a {\it different} $U(N)$ group. For tensors with $2d$ indices we have the transformation law
\begin{eqnarray}
 \Phi_{j_1\dots j_d}^{i_1\dots i_d}(x)'&=&(g_1)_{j_1}^{k_1} \cdots (g_d)_{j_d}^{k_d} (g_{\bar{1}}^\dagger)_{l_1}^{i_1} \cdots (g^\dagger_{\bar{d}})_{l_d}^{i_d} \Phi_{k_1\dots k_d}^{l_1\dots l_d}(x),\label{unitaryaction}\\
 \overline{\Phi}^{j_1\dots j_d}_{i_1\dots i_d}(x)'&=&(g^\dagger_1)^{j_1}_{k_1} \cdots (g^\dagger_d)^{j_d}_{k_d} (g_{\bar{1}})^{l_1}_{i_1} \cdots (g_{\bar{d}})^{l_d}_{i_d} \overline{\Phi}^{k_1\dots k_d}_{l_1\dots l_d}(x),\label{unitaryactionconjugate}
\end{eqnarray}
with $(g_1,\dots,g_d,g_{\bar{1}},\dots,g_{\bar{d}})$ an element of $G_{d\bar{d}}(N)$. Again, the contraction $\Phi\cdot \overline{\Phi}=\Phi^{i_1\dots i_d}_{j_1\dots j_d}\overline{\Phi}_{i_1\dots i_d}^{j_1\dots j_d}$ is invariant under the action of the group.\\

Now, suppose that the initial tensor theory is globally invariant under $G_{d\bar{d}}(N)$. This means that the Lagrangian, build on tensors $\Phi$ and $\overline{\Phi}$ which transform as \eqref{unitaryaction} and \eqref{unitaryactionconjugate},
 is invariant under $G_{d\bar{d}}(N)$.
 That is
\begin{equation}
\mathcal{L}\big[\Phi(x)\big]=\mathcal{L}\big[\Phi'(x)\big].
\end{equation}
Apart from the invariance under the gauge group $G_{d\bar{d}}(N)$, the details of the Lagrangian are irrelevant in the present discussion. We will assume that whatever the Lagrangian is, it allows SSB. We will apply the same methodology as the scalar case, treated in section \ref{scalar0}.

The path integral of the tensor theory is written as
\begin{multline}\label{symPI}
Z_{\epsilon}\big[J,\overline{J}\big]=\frac{1}{\mathcal{N}}\int D\Phi D\overline{\Phi}\exp\bigg(\text{i}\int d^4x\Big\{\mathcal{L}\big[\Phi(x)\big] \\+ \overline{J}(x)\cdot \Phi(x)+J(x)\cdot \overline{\Phi}(x) +\text{i}\epsilon|\Phi(x)-v|^2  \Big \}   \bigg),
\end{multline}
 with 
\begin{equation}
\mathcal{N}=Z_{\epsilon}\big[0,0\big],
\end{equation}
and 
\begin{equation}
|\Phi(x)-v|^2=\Phi(x)\cdot\overline{\Phi}(x)-\Phi(x)\cdot\overline{v}-v\cdot\overline{\Phi}(x)+v\cdot\overline{v},
\end{equation}
where the dot indicates full contraction of indices,
\begin{equation}
X\cdot \overline{Y}=X_{i_1\dots i_d}^{j_1 \dots j_d}\overline{Y}^{i_1\dots i_d}_{j_1\dots j_d}.
\end{equation}
The interpretation of (\ref{symPI}) is standard, with $J^{i_1\dots i_d}_{j_1 \dots j_d}(x)$ and $\overline{J}^{i_1\dots i_d}_{j_1\dots j_d}(x)$ the external source fields which are used to extract the Green functions from   (\ref{symPI}) via derivation. For $\epsilon=0$ the theory enjoys the full symmetry $G_{d\bar{d}}(N)$. For $\epsilon \neq 0$, the term $|\Phi(x)-v|^2 $ is chosen so as to break the $G_{d\bar{d}}(N)$-symmetry, to induce SSB. After the calculation we take the limit $\epsilon\to 0$. The tensor $v_{j_1\dots j_d}^{i_1\dots i_d}$ is not spacetime-dependent. It may be thought as the boundary value of $\Phi(x)$ at infinity. When SSB occurs, the different degenerate vacua that the theory presents correspond to different choices of $v$. The effect of the $\epsilon$-term is to pick a particular vacuum of the theory. As said above, we are assuming that the Lagrangian presents SSB, which in this discussion means that the field configurations allow a non-zero $v$ value at infinity.   \\

In this paper, we will be interested in $G_{d\bar{d}}(N)$-invariant theories that spontaneously break into different patterns.   The different choices of $v$ and their relation with the several patterns of symmetry breaking will be clarified in the next section.

\section{Symmetry breaking patterns}\label{SSpatterns}
Different choices of $v$ in the $\epsilon$-term lead to different symmetry breaking patterns. For instance, if $v=0$ there is not symmetry breaking whereas for a generic $v$ the whole group $G_{d\bar{d}}(N)$ gets spontaneously broken and there is no remaining continuous symmetry\footnote{In more general SSB with generic $v$, some discrete symmetries might remain.}.
We will explore the choices of $v$ that induce SSB into diagonal subgroups of $G_{d\bar{d}}(N)$. The general idea is that $v$ must be an invariant of the subgroup we wish to break into. In this paper we are studying the patterns 
\be \label{SSBdiag}
G_{d\bar{d}}(N)=\prod_{k=1}^d[U_k(N)\times U_{\bar k}(N)]\longrightarrow G_\omega(N)=\prod_{\alpha=1}^\omega\text{Diag}[H_{\alpha}],
\ee
where each $H_{\alpha}$ is a tensor product of $2n_\alpha$ different unitary groups  ($n_\alpha$ barred and $n_\alpha$ unbarred) such that 
\be\label{interH}
H_{\alpha}\cap H_\beta=\emptyset,\qquad \alpha\neq \beta,\qquad \sum_{\alpha=1}^\omega (2n_\alpha)=2d.
\ee
Be aware that each SSB pattern is associated with a partition of $d$ elements with $\omega$ parts, $\sum_{\alpha=1}^\omega n_\alpha=d$. Notice, as well, that $\text{Diag}[H_{\alpha}]$ is a unitary group and it will often be denoted $U_\alpha(N)$ along this paper.
The SSB into the full diagonal group $G_1(N)=U(N)$ corresponds to $\omega=1$, whereas for $\omega=d$ the remaining symmetry group is $G_d(N)=\prod_{\alpha=1}^dU_\alpha(N)$. 

The number of Goldstone boson that will result after a symmetry breaking $G\to H$ is counted by subtracting the number of generators of the initial and remaining symmetry groups, $d_G-d_H$. For the SSB pattern \eqref{SSBdiag}, we have
\be\label{conteo}
d_{G_{d\bar{d}}(N)} -d_{G_{\omega}(N)} =2dN^2-\omega N^2=(2d-\omega)N^2.
\ee
For the choices \eqref{SSBdiag} the invariants  under the corresponding subgroup can be constructed using Kronecker deltas. Thus, in general,
\be\label{vsigma}
v^{i_1\dots i_d}_{j_1\dots j_d}=\sum_{\sigma\in S_d}v_\sigma\delta^{i_1\dots i_d}_{j_{\sigma(1)}\dots j_{\sigma(d)}},
\ee
will induce SSB into any $G_\omega(N)$ for suitable $v_\sigma$ parameters, where
\be
\delta^{i_1\dots i_d}_{j_1\dots j_{d}}=\delta^{i_1}_{j_1}\cdots \delta^{i_d}_{j_d}.
\ee
To make the former statement clearer we will examine three examples; the case $d=1$, for which  $v^i_j\propto \delta^i_j$, and the case $d=2$, for which  $v^{i_1i_2}_{j_1j_2}\propto \delta^{i_1i_2}_{j_1j_2}$ or $v^{i_1i_2}_{j_1j_2}\propto \delta^{i_1i_2}_{j_2j_1}$.
In the path integral, the $\epsilon$-term is of the form $\epsilon v\Phi$. We wish to know which subgroup of $G_{d\bar{d}}(N)$ the $\epsilon$-term is invariant for a given $v$.  

For $d=1$, the tensor $\Phi$ transforms under $G_{1\bar{1}}(N)=U_1(N)\times U_{\bar{1}}(N)$. The $\epsilon$-term is $\epsilon v^i_j\Phi^j_i\propto \delta^i_j\Phi^j_i$. It is clear in this simple case that the $\epsilon$-term, hence the path integral, is only invariant under the diagonal $U(N)$ subgroup of $G_{1\bar{1}}(N)$, i.e., 
\be
\delta^i_j\Phi^{\prime j}_i=\delta^i_jg^{\dagger k}_ig^j_l\Phi_k^l=\delta^k_l\Phi^{l}_k.
\ee
This tells us that the choice $v^i_j\propto \delta^i_j$ induce SSB into the diagonal subgroup of $G_{1\bar{1}}(N)$. We would like to stress that the role of $\delta$ is to link two indices, up and downstairs, which results in the identification of $U_1(N)$ and $U_{\bar{1}}(N)$.

In the case of $d=2$, the choice $v^{i_1i_2}_{j_1j_2}\propto \delta^{i_1i_2}_{j_1j_2}$ will make the $\epsilon$-term invariant under the group 
\be\label{G2id}
G_2(N)=\text{Diag}[U_1(N)\times U_{\bar{1}}(N)]\times \text{Diag}[U_2(N)\times U_{\bar{2}}(N)].  
\ee
Using the notation \eqref{genericgdd}, a general element of $G_2(N)$ is $(g,h,g,h)$.
Whereas for $v^{i_1i_2}_{j_1j_2}\propto \delta^{i_1i_2}_{j_2j_1}$, the $\epsilon$-term will be invariant under
\be\label{G212}
G'_2(N)=\text{Diag}[U_1(N)\times U_{\bar{2}}(N)]\times \text{Diag}[U_2(N)\times U_{\bar{1}}(N)],  
\ee
where a general element of $G'_2(N)$ can be written as $(g,h,h,g)$.

The next step is to consider the linear combination $v^{i_1i_2}_{j_1j_2}=v_1\delta^{i_1i_2}_{j_1j_2}+v_2\delta^{i_1i_2}_{j_2j_1}$. The $\epsilon$-term will be invariant under the intersection $G_2(N)\cap G'_2(N)$, whose general element is $(g,g,g,g)$, that is, an element of $G_1(N)$.

In terms of $v_\sigma$ the simplest cases to study are breaking patterns into $G_d(N)$.
There are many different $v$'s which make that job, one for each monomial $\delta^{i_1\dots i_d}_{j_{\sigma(1)}\dots j_{\sigma(d)}}$. For $N\geq d$ there is one monomial per permutation $\sigma \in S_d$. So and overall of $d!$ monomials and, therefore, $d!$ different patterns. If $N<d$ then not all of them are linearly independent and the number of SSB patterns into $G_d(N)$ are counted by the formula
\be\label{exactmon}
 \text{Number of linearly independent monomials}=\sum_{\substack{R\vdash d\\
l(R)\leq N}} d_R^2,
\ee
where $d_R$ is the dimension of the irrep\footnote{Note that the irreducible representations (irreps) of $S_d$ are labeled by Young diagrams with $d$ boxes.} $R$ of  $S_d$, $R\vdash d$ tells that the Young diagram $R$ has $n$ boxes and $l(R)$ is the number of rows of $R$. Formula \eqref{exactmon} counts the number of linearly independent monomials for any $N$. However, the interest is usually focused on large $N$ theories. In these cases, the condition $l(R)\leq N$ is  fulfilled and the sum \eqref{exactmon} is always  $d!$.
 
If we wish to induce SSB into any other subgroup of the type $G_\omega(N)$ we need to consider a sum of monomials as indicated in \eqref{vsigma}. We shall see that there is a minimal choice of $v$  for any symmetry breaking pattern. Specifically, and this is one of the main results of the paper, we are showing in the next sections that $v$ built on just two summands in \eqref{vsigma}, 
\be\label{2summands}
v^{i_1\dots i_d}_{j_1\dots j_d}=v_\sigma\delta^{i_1\dots i_d}_{j_{\sigma(1)}\dots j_{\sigma(d)}}+v_{\sigma'}\delta^{i_1\dots i_d}_{j_{\sigma'(1)}\dots j_{\sigma'(d)}}, 
\ee
appropriately chosen, induces any pattern of SSB in \eqref{SSBdiag}. For this purpose we will present a visual diagrammatic  correspondence between the choices of $v$ and the SSB patterns.

\subsection{Diagrams}
As commented above, the role of delta monomials is to identify up and downstairs groups. In general, for one monomial, the SSB pattern is given by
\be\label{monoassociation}
\delta^{i_1\dots i_d}_{j_{\sigma(1)}\dots j_{\sigma(d)}}\longrightarrow \prod_{k=1}^d\text{Diag}[U_k(N)\times U_{\sigma(\bar{k})}(N)].
\ee
This suggests the following diagrammatical correspondence
\be\label{simplegeneral}
\delta^{i_1\dots i_d}_{j_{\sigma(1)}\dots j_{\sigma(d)}}\longrightarrow \begin{array}{ccc}
g_1&\dots& g_d\\
|&\dots&|\\
g_{\sigma(\bar{1})}&\dots & g_{\sigma(\bar{d})}
\end{array}.
\ee   
For $d=2$, the two deltas that yield to the {\it different} symmetry breaking patterns shown in \eqref{G2id} and \eqref{G212} can be mapped into the diagrams
 \be \label{simpledeltas}
\delta^{i_1i_2}_{j_1j_2}\longrightarrow 
\begin{array}{cc}
~g_1~&~g_2~\\
~|~&~|~\\
~g_{\bar{1}}~&~g_{\bar{2}}~
\end{array}
\qquad ,
\qquad
\delta^{i_1i_2}_{j_2j_1}\longrightarrow \begin{array}{cc}
~g_1~&~g_2~\\
~|~&~|~\\
~g_{\bar{2}}~&~g_{\bar{1}}~
\end{array}.
\ee    
As said before, the linear combination of deltas comes along with the intersection of the groups that each delta induces.
Precisely, it is when considering intersections where the diagrams will show their usefulness. 

Diagrams representing intersections, which will be called ``intersection-diagrams,''  are constructed by concatenation of two diagrams of the type  \eqref{simplegeneral}, and then joining the equal  elements upstairs and their ``bar'' counterparts downstairs. This is exemplified in Fig.\ref{simple1}, which represents the intersection of two monomials for $d=2$.
\setlength{\unitlength}{1cm}
\begin{figure}[ht]
\begin{center}
\begin{picture}(10,5)
\put(1,2){
$v_1\delta^{i_1i_2}_{j_{1}j_{2}}+v_2\delta^{i_1i_2}_{j_2j_1}~~\longrightarrow ~~\begin{array}{cccc}
        ~g_1~ & ~g_2 ~& ~g_1 ~& ~g_2 ~ \\
        ~| ~& ~| ~& ~| ~&~ | ~ \\
        ~g_{\bar{1}}~ & ~g_{\bar{2}} ~& ~g_{\bar{2}} ~& ~g_{\bar{1}}~
     \end{array}.$} 

   {\color{blue}
   \put(5.65,3.25){\line(1,0){1.55}}
  \put(5.65,3.25){\line(0,-1){.35}}
   \put(7.2,3.25){\line(0,-1){.35}}}
  
  {\color{blue}
  \put(6.3,3.4){\line(1,0){1.55}}
  \put(6.3,3.4){\line(0,-1){.5}}
   \put(7.85,3.4){\line(0,-1){.5}}}
   
    {\color{blue}
   \put(5.3,.8){\line(1,0){2.35}}
  \put(5.3,.8){\line(0,1){.4}}
   \put(7.65,.8){\line(0,1){.4}}}
  
  {\color{blue}
  \put(6,.9){\line(1,0){.75}}
  \put(6,.9){\line(0,1){.35}}
   \put(6.75,.9){\line(0,1){.35}}}
   
\end{picture}
\caption[]{Intersection-diagram for $d=2$ with $\sigma=(1)(2)$ and $\sigma'=(12)$.} \label{simple1}
\end{center}
\end{figure}

The diagram of  Fig.\ref{simple1} displays just one cycle, which is in correspondence with the only $U(N)$ group that results from $G_2(N)\cap G'_2(N)$. This is not a coincidence. It turns out that the cycle structure of the intersection-diagrams\footnote{We will refer as the cycle structure  the set of all loops, together with their length, that fully connect the diagram. Length is the number of elements the loop involves. Cycle structures are in one-to-one correspondence with partitions of $d$.} corresponds  with the different symmetry breaking patterns. Therefore, the number of cycles of the diagram will be $\omega$ in the remaining group $G_\omega(N)$. This statement will be expanded in the subsequent paragraphs.

As a consistency check of the diagrammatic correspondence, let us see how it works when $v$ is just one monomial, say, $v^{i_1i_2}_{j_1j_2}=v_1\delta^{i_1i_2}_{j_1j_2}$. This choice breaks into the group $G_2(N)$ in the equation \eqref{G2id}, which corresponds to the first diagram of \eqref{simpledeltas}. 
Trivially, $v$ may be written as the sum $\frac{1}{2}(v_1\delta^{i_1i_2}_{j_1j_2}+   v_1\delta^{i_1i_2}_{j_1j_2})$. On the one hand, we have two (same) monomials. In this case the sum is interpreted as a SSB term to the group $G_2(N)\cap G_2(N)$. On the other hand, using the rules of the intersection discussed above, we get the diagram in Fig.\ref{simple2}, 
\setlength{\unitlength}{1cm}
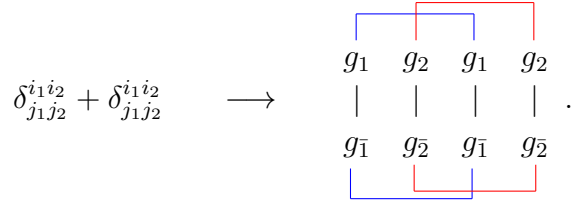
\begin{figure}[ht]
\begin{center}
\begin{picture}(10,5)
\put(1,2){
$\delta^{i_1i_2}_{j_{1}j_{2}}+\delta^{i_1i_2}_{j_1j_2}~~~~~\longrightarrow~~~~\begin{array}{cccc}
        ~g_1~ & ~g_2 ~& ~g_1 ~& ~g_2 ~ \\
        ~| ~& ~| ~& ~| ~&~ | ~ \\
        ~g_{\bar{1}}~ & ~g_{\bar{2}} ~& ~g_{\bar{1}} ~& ~g_{\bar{2}}~
     \end{array}.$} 

   {\color{blue}
   \put(5.65,3.25){\line(1,0){1.55}}
  \put(5.65,3.25){\line(0,-1){.35}}
   \put(7.2,3.25){\line(0,-1){.35}}}
  
  {\color{red}
  \put(6.3,3.4){\line(1,0){1.55}}
  \put(6.3,3.4){\line(0,-1){.5}}
   \put(7.85,3.4){\line(0,-1){.5}}}
   
    {\color{blue}
   \put(5.3,.8){\line(1,0){1.6}}
  \put(5.3,.8){\line(0,1){.4}}
   \put(6.9,.8){\line(0,1){.4}}}
  
  {\color{red}
  \put(6,.9){\line(1,0){1.6}}
  \put(6,.9){\line(0,1){.35}}
   \put(7.6,.9){\line(0,1){.35}}}
    
\end{picture}
\caption[]{Intersection-diagram of $G_2(N)\cap G_2(N)$.} \label{simple2}
\end{center}
\end{figure}
 which is a diagram with two cycles. This tells us that the remaining symmetry group is $G_2(N)$, in perfect agreement with $G_2(N)\cap G_2(N)=G_2(N)$.

Let us discuss the case $d=3$. As we show in Fig.\ref{double2},  for different choices of $\sigma$ and $\sigma'$ the diagram will display different cycle structures,  which are not shown in Fig.\ref{double2} because $\sigma$ and $\sigma'$ have not been specified yet. We will explore some explicit examples in the following.
\setlength{\unitlength}{1cm}
\begin{figure}[ht]
\begin{picture}(4,4)
\put(1,2){
$v_1\delta^{i_1i_2i_3}_{j_{\sigma(1)}j_{\sigma(2)}j_{\sigma(3)}}+v_2\delta^{i_1i_2i_3}_{j_{\sigma'(1)}j_{\sigma'(2)}j_{\sigma'(3)}}\longrightarrow \begin{array}{cccccc}
        g_1 & g_2 & g_3 & g_1 & g_2 & g_3  \\
        | & | & | & | & | & | \\
        g_{\sigma(\bar{1})} & g_{\sigma(\bar{2})} & g_{\sigma(\bar{3})} & g_{\sigma'(\bar{1})} & g_{\sigma'(\bar{2})} & g_{\sigma'(\bar{3})} 
      \end{array}.$}
 \put(8.2,3.25){\line(1,0){2.8}}
  \put(8.2,3.25){\line(0,-1){.35}}
   \put(11,3.25){\line(0,-1){.35}}
  
   \put(9.1,3.4){\line(1,0){2.9}}
  \put(9.1,3.4){\line(0,-1){.5}}
   \put(12,3.4){\line(0,-1){.5}} 
    
   \put(10,3.55){\line(1,0){3}}
  \put(10,3.55){\line(0,-1){.65}}
   \put(13,3.55){\line(0,-1){.65}}   
   
\end{picture}
\caption[]{Intersection-diagram for $d=3$. Different choices of $\sigma$ and $\sigma'$ will lead to different cycle structures.} \label{double2}
\end{figure}

In Fig.\ref{double3} we show the diagram of $d=3$ where $\sigma$ is the identity and $\sigma'$ is a transposition. The identity induces the SSB into 
\be\label{G3id}
G_3(N)=\text{Diag}[U_1(N)\times U_{\bar{1}}(N)]\times \text{Diag}[U_2(N)\times U_{\bar{2}}(N)]\times\text{Diag}[U_3(N)\times U_{\bar{3}}(N)].  
\ee
Whereas $\sigma'=(12)(3)$ induces the SSB into
\be\label{G312}
G'_3(N)=\text{Diag}[U_1(N)\times U_{\bar{2}}(N)]\times \text{Diag}[U_2(N)\times U_{\bar{1}}(N)]\times\text{Diag}[U_3(N)\times U_{\bar{3}}(N)].  
\ee
Now, 
\begin{multline}\label{G3}
G_3(N)\cap G'_3(N)= \\
\text{Diag}[U_1(N)\times U_{\bar{1}}(N)\times U_2(N)\times U_{\bar{2}}(N)]\times \text{Diag}[U_3(N)\times U_{\bar{3}}(N)] \\
=U(N)\times U'(N)=G_2(N).
\end{multline}
A detailed description of the symmetry breaking pattern is written in the second line of equation \eqref{G3}. Note that this information is encoded in the diagram in Fig.\ref{double3} and can be easily read off.
\setlength{\unitlength}{1cm}
\begin{figure}[ht]
\begin{center}
\begin{picture}(10,5)
\put(1,2){
$v_1\delta^{i_1i_2i_3}_{j_{1}j_{2}j_{3}}+v_2\delta^{i_1i_2i_3}_{j_2j_1j_3}\longrightarrow \begin{array}{cccccc}
        ~g_1~ & ~g_2 ~& ~g_3 ~& ~g_1 ~& ~g_2 ~&~ g_3  \\
        ~| ~& ~| ~& ~| ~&~ | ~& ~|~ &~ | \\
        ~g_{\bar{1}}~ & ~g_{\bar{2}} ~&~g_{\bar{3}} ~& ~g_{\bar{2}} ~& ~g_{\bar{1}}~ & ~g_{\bar{3}} 
     \end{array}.$} 

   {\color{blue}
   \put(5.65,3.25){\line(1,0){2.35}}
  \put(5.65,3.25){\line(0,-1){.35}}
   \put(8,3.25){\line(0,-1){.35}}}
  
  {\color{blue}
  \put(6.3,3.4){\line(1,0){2.35}}
  \put(6.3,3.4){\line(0,-1){.5}}
   \put(8.65,3.4){\line(0,-1){.5}}}
    
  {\color{red}\put(7,3.55){\line(1,0){2.35}}
  \put(7,3.55){\line(0,-1){.65}}
   \put(9.35,3.55){\line(0,-1){.65}}} 
   
    {\color{blue}
   \put(5.3,.8){\line(1,0){3.15}}
  \put(5.3,.8){\line(0,1){.4}}
   \put(8.45,.8){\line(0,1){.4}}}
  
  {\color{blue}
  \put(5.9,.9){\line(1,0){1.65}}
  \put(5.9,.9){\line(0,1){.35}}
   \put(7.55,.9){\line(0,1){.35}}}
    
  {\color{red}\put(6.6,.6){\line(1,0){2.35}}
  \put(6.6,.6){\line(0,1){.65}}
   \put(8.95,.6){\line(0,1){.65}}} 
   
\end{picture}
\caption[]{Diagrammatic representation of two deltas, with $\sigma=(1)(2)(3)$ and $\sigma'=(12)(3)$.} \label{double3}
\end{center}
\end{figure}

As a last example, we will regard the $d=3$ with $\sigma=(1)(23)$ and $\sigma'=(12)(3)$.
The monomial with the permutation $\sigma$ induces SSB into
\be\label{G3id}
G_3(N)=\text{Diag}[U_1(N)\times U_{\bar{1}}(N)]\times \text{Diag}[U_2(N)\times U_{\bar{3}}(N)]\times\text{Diag}[U_3(N)\times U_{\bar{2}}(N)].  
\ee
Whereas the monomial with $\sigma'=(12)(3)$ leads to
\be\label{G312}
G'_3(N)=\text{Diag}[U_1(N)\times U_{\bar{2}}(N)]\times \text{Diag}[U_2(N)\times U_{\bar{1}}(N)]\times\text{Diag}[U_3(N)\times U_{\bar{3}}(N)].  
\ee
Now, 
\begin{multline}\label{G3prime}
G_3(N)\cap G'_3(N)= \\
\text{Diag}[U_1(N)\times U_{\bar{1}}(N)\times U_2(N)\times U_{\bar{2}}(N)\times U_3(N)\times U_{\bar{3}}(N)] \\
=U(N)=G_1(N).
\end{multline}
For this case the SSB into the diagonal group is in perfect agreement with the diagram in Fig.\ref{double4} which shows only one cycle.
\setlength{\unitlength}{1cm}
\begin{figure}[ht]
\begin{center}
\begin{picture}(10,5)
\put(1,2){
$v_1\delta^{i_1i_2i_3}_{j_{1}j_{2}j_{3}}+v_2\delta^{i_1i_2i_3}_{j_2j_1j_3}\longrightarrow \begin{array}{cccccc}
        ~g_1~ & ~g_2 ~& ~g_3 ~& ~g_1 ~& ~g_2 ~&~ g_3  \\
        ~| ~& ~| ~& ~| ~&~ | ~& ~|~ &~ | \\
        ~g_{\bar{1}}~ & ~g_{\bar{3}} ~&~g_{\bar{2}} ~& ~g_{\bar{2}} ~& ~g_{\bar{1}}~ & ~g_{\bar{3}} 
     \end{array}.$} 

   {\color{blue}
   \put(5.65,3.25){\line(1,0){2.35}}
  \put(5.65,3.25){\line(0,-1){.35}}
   \put(8,3.25){\line(0,-1){.35}}}
  
  {\color{blue}
  \put(6.3,3.4){\line(1,0){2.35}}
  \put(6.3,3.4){\line(0,-1){.5}}
   \put(8.65,3.4){\line(0,-1){.5}}}
    
  {\color{blue}\put(7,3.55){\line(1,0){2.35}}
  \put(7,3.55){\line(0,-1){.65}}
   \put(9.35,3.55){\line(0,-1){.65}}} 
   
    {\color{blue}
   \put(5.3,.8){\line(1,0){3.15}}
  \put(5.3,.8){\line(0,1){.4}}
   \put(8.45,.8){\line(0,1){.4}}}
  
  {\color{blue}
  \put(6.7,.9){\line(1,0){.77}}
  \put(6.7,.9){\line(0,1){.35}}
   \put(7.47,.9){\line(0,1){.35}}}
    
  {\color{blue}\put(5.8,.6){\line(1,0){3.15}}
  \put(5.8,.6){\line(0,1){.65}}
   \put(8.95,.6){\line(0,1){.65}}} 
   
\end{picture}
\caption[]{Intersection-diagram corresponding to $\sigma=(1)(23)$ and $\sigma'=(12)(3)$. It displays only one cycle, which indicates that it breaks into the diagonal group.} \label{double4}
\end{center}
\end{figure}
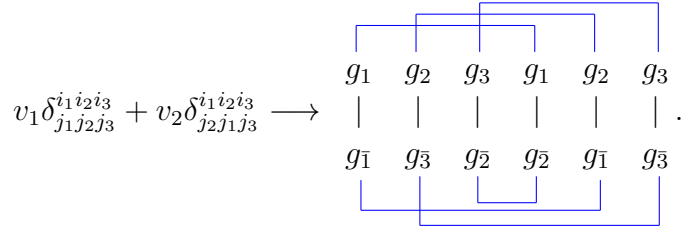

With these examples we have shown in detail how a particular SSB pattern is associated with an intersection-diagram. Moreover, for $G_\omega(N)$,
\be
\omega= \text{number of cycles of the intersection-diagram.}
\ee 
\subsection{Parameter space of SSB}\label{MS}
Now, using the intersection-diagrams we can  prove the statement leading to \eqref{2summands}. That is, {\it there exists a minimal choice of $v$  for any symmetry breaking pattern, namely, two summands}.  Be aware that because two deltas fix uniquely the intersection-diagram, in order to complete the proof it is sufficient to see that for an arbitrary SSB pattern there is always (at least) one intersection-diagram associated to it.

Note that any SSB can be associated to a partition of $d$, given by the $n_\alpha$ values. However, the choice of a partition does not specify completely the SSB pattern.

In order to associate an intersection-diagram to \eqref{SSBdiag} we should proceed as follows. First, we draw a plain (with no numbers) diagram 
\be\label{plain}
\underbrace{
 \begin{array}{ccc|ccc}
g&\dots& g&g&\dots& g\\
|&\dots&|&|&\dots&|\\
g&\dots & g&g&\dots & g
\end{array}}_{2d \text{ slots}}.
\ee   
Second, we draw in the diagram the cycle structure associated to a given SSB pattern, which can be read off from the set $\{H_{\alpha}\}$. For each of the $\omega$ $H_{\alpha}$ groups  we join $n_\alpha$ slots from the LHS of \eqref{plain} with $n_\alpha$ slots of the RHS of \eqref{plain} in a single loop. This turns \eqref{plain} into a plain cycle-structured diagram with closed loops. We would like to emphasize that such procedure always fully connect the plain diagram. This happens because $n_1+n_2+\dots +n_\omega=d$ is a partition of $d$, and the groups $H_{\alpha}$ do not intersect each other, as stated in \eqref{interH}.

The next step is to complete the diagram by writing the subscript of each element according to the given SSB pattern. For this purpose   we focus on the cycle associated to $H_{\alpha}$  and write, as subscripts, the labels of the different unitary groups $H_{\alpha}$ contains. For instance, if $n_\alpha=2$, we have the generic group
\be 
H_{\alpha}=U_a\times U_{\bar{b}}\times U_c\times U_{\bar{d}}.
\ee
Then the labelling process on the cycle of the diagram corresponding to $H_{\alpha}$ can be chosen as
\setlength{\unitlength}{1cm}
\begin{figure}[H]\label{cyclelabel}
\begin{center}
\begin{picture}(10,1.5)
$
 \begin{array}{cc|cc}
~~g~~& ~~g~~&~~g~~& ~~g~~\\
|&|&|&|\\
g& g&g& g
\end{array}\longrightarrow
\begin{array}{cc|cc}
~~g_a~~&~~ g_c~~&~~g_c~~&~~ g_a~~\\
|&|&|&|\\
g_{\bar{d}}& g_{\bar{b}}&g_{\bar{d}}& g_{\bar{b}}
\end{array}.
$  
  {\color{blue}
   \put(-8.5,1.5){\line(1,0){2.75}}
  \put(-8.5,1.5){\line(0,-1){.55}}
   \put(-5.75,1.5){\line(0,-1){.55}}}

  {\color{blue}
   \put(-7.75,1.4){\line(1,0){.9}}
  \put(-7.75,1.4){\line(0,-1){.45}}
   \put(-6.85,1.4){\line(0,-1){.45}}}

 {\color{blue}
   \put(-8.8,-1){\line(1,0){1.75}}
  \put(-8.8,-1){\line(0,1){.2}}
   \put(-7.05,-1){\line(0,1){.2}}}

  {\color{blue}
   \put(-8.05,-1.2){\line(1,0){1.8}}
  \put(-8.05,-1.2){\line(0,1){.45}}
   \put(-6.25,-1.2){\line(0,1){.45}}}

 {\color{blue}
   \put(-4.5,1.5){\line(1,0){3.1}}
  \put(-4.5,1.5){\line(0,-1){.55}}
   \put(-1.4,1.5){\line(0,-1){.55}}}

  {\color{blue}
   \put(-3.55,1.4){\line(1,0){1}}
  \put(-3.55,1.4){\line(0,-1){.45}}
   \put(-2.55,1.4){\line(0,-1){.45}}}

 {\color{blue}
   \put(-4.7,-1){\line(1,0){2}}
  \put(-4.7,-1){\line(0,1){.2}}
   \put(-2.7,-1){\line(0,1){.2}}}

  {\color{blue}
   \put(-3.85,-1.2){\line(1,0){2.05}}
  \put(-3.85,-1.2){\line(0,1){.45}}
   \put(-1.8,-1.2){\line(0,1){.45}}}

\end{picture}
\end{center}
\end{figure}
\vspace{1cm}
A complete intersection-diagram is obtained by applying the same labelling prescription to each $H_{\alpha}$.
Now, the two delta monomials \eqref{2summands} can be read off from the full diagram. This concludes the proof.

As a remark, notice that the process of labelling is not unique. In general, there will be several pairs of deltas that would induce the same SSB pattern. This fact does not affect the conclusion of our argument since the aim is to show that any SSB pattern can be induced only with two parameters, as in  \eqref{2summands}.

As a second remark, from the two permutations $\sigma$ and $\sigma'$ in \eqref{2summands}, there is a straight way of reading the partition associated to a SSB pattern, which is a valuable information of the SSB pattern.  As seen in this section, the partition of the SSB pattern is in correspondence with the cycle structure of the intersection-diagram. 

Now, it is interesting to read the intersection-diagrams as a composition of permutations. To make this point clearer let us picture the composition of  two permutations that we used in the examples in Fig.\ref{double3} and Fig.\ref{double4}  
\be\label{compo}
\sigma'^{-1} \cdot\sigma=
\begin{array}{c}
\left(  \begin{array}{ccc}
2&1&3\\
1&2&3
\end{array}\right)\\
\begin{array}{ccc}
\big{|}&\big{|}&\big{|}
\end{array}\\
\left(  \begin{array}{ccc}
1&2&3\\
1&2&3
\end{array}\right)
\end{array}=
\left(  \begin{array}{ccc}
2&1&3\\
1&2&3
\end{array}\right)\quad ,\quad 
\sigma'^{-1} \cdot \sigma=
\begin{array}{c}
\left(  \begin{array}{ccc}
2&1&3\\
1&2&3
\end{array}\right)\\
\begin{array}{ccc}
\big{|}&\big{|}&\big{|}
\end{array}\\
\left(  \begin{array}{ccc}
1&2&3\\
1&3&2
\end{array}\right)
\end{array}=
\left(  \begin{array}{ccc}
2&1&3\\
1&3&2
\end{array}\right),
\ee
respectively. Note that the compositions pictured in \eqref{compo} can be traced in the diagrams. The composition is realized as we follow a path starting on the upper right side of the diagram,  move downstairs, then towards the left side and ending on the upper left side of the diagram.
From this observation we conclude  that the cycle structure of the diagram coincides with the cycle structure of the permutation $\sigma'^{-1}\cdot\sigma$. Recall that the cycle structure of a permutation is indeed a partition \cite{fulton}.

\section{Matrix fields as Goldstone bosons}\label{general formalism}
The next step in our discussion is to identify the Goldstone bosons associated to a certain SSB pattern which, as discussed in the previous section, is induced  by a non-vanishing $v$. To this aim we will derive the so-called  WT identities which are identities among Green functions that arise from the path integral (\ref{symPI}) with the $\epsilon$-term above defined. This will be a generalization of the method exposed in section \ref{scalar0}. 
\subsection{Identities from the symmetry of the path integral}
First, notice that path integral  (\ref{symPI}) is invariant under the change of variables $\Phi(x)\to \Phi'(x)$ and $\overline{\Phi}(x)\to \overline{\Phi}'(x)$ as in (\ref{unitaryaction}) and (\ref{unitaryactionconjugate}), respectively. 
Let us write the elements of the special unitary group in exponential form as
\begin{equation}\label{expmap}
g_k=e^{\text{i}\theta_a^{k} T_a}, \qquad g_{\bar{k}}=e^{\text{i}\theta_a^{\bar{k}} T_a},
\end{equation}
where $T_a$ are the generators of the algebra  $\mathfrak{u}(N)$. The labels $k$ and $\bar{k}$   have been added to indicate the slot in the tensor indices the element $g_k$ is acting on. The algebra $\mathfrak{u}(N)$ is generated by $t_a$ which denote the generators of  $\mathfrak{su}(N)$ plus the identity, i.e., $T_a=(\frac{1}{\sqrt{2N}}\mathds{1},t_a)$. We will also choose the generators normalized as $\text{Tr}(T_aT_b)=\frac{1}{2}\delta_{ab}$. 

We now proceed to apply the invariance of the path integral under the field change. 
With the parametrization (\ref{expmap}), it is expressed as 
\begin{equation}\label{derivatingZ}
\frac{\partial Z(J(x),\overline{J}(x))}{\partial \theta_a^{k}}=\frac{\partial Z(J(x),\overline{J}(x))}{\partial \theta_a^{\bar{k}}}=0,
\end{equation}
where $k$ and $\bar{k}$ refer to the $2d$ transformations
\begin{eqnarray}\label{costaction}
\Phi^{i_1\dots i_d}_{j_1\dots j_d}(x)&\longrightarrow& (g_k)_{j}^{i_k}\Phi^{i_1\dots j \dots i_d}_{j_1\dots j_d}(x),\quad k=1,\dots, d,\nonumber \\
\Phi^{i_1\dots i_d}_{j_1\dots j_d}(x)&\longrightarrow& (g^\dagger_{\bar{k}})^{j}_{j_{\bar{k}}}\Phi^{i_1\dots i_d}_{j_1\dots j \dots j_d}(x),\quad \bar{k}=1,\dots, d \ ,
\end{eqnarray}
which lead, for $\theta_a^k$, to the collection of identities
\begin{equation}\label{basicidentities}
\int d^4x \langle \overline{J}(x)T_a^{(k)}\Phi(x)-\overline{\Phi}(x)T_a^{(k)}J(x)-\text{i}\epsilon\big(\overline{v}T_a^{(k)}\Phi(x)-\overline{\Phi}(x)T_a^{(k)}v\big)\rangle_{J,\epsilon}=0.
\end{equation}
and, derivating with respect to $\theta_a^{\bar{k}}$ leads to
\begin{equation}\label{basicidentities2}
\int d^4x \langle \Phi(x)T_a^{(\bar{k})}\overline{J}(x)-J(x)T_a^{(\bar{k})}\overline{\Phi}(x)-\text{i}\epsilon\big(\Phi(x)T_a^{(\bar{k})}\overline{v}-vT_a^{(\bar{k})}\overline{\Phi}(x)\big)\rangle_{J,\epsilon}=0.
\ee
We have used the notation
\begin{multline}\label{Fdefinition}
\langle F(\Phi)\rangle_{J,\epsilon}=\frac{1}{\mathcal{N}}\int D\Phi D\overline{\Phi}F(\Phi) \exp\bigg(\text{i}\int d^4x\Big\{\mathcal{L}\big(\Phi(x)\big)\\ + \overline{J}(x)\cdot \Phi(x)+J(x)\cdot\overline{\Phi}(x) +\text{i}\epsilon |\Phi(x)-v|^2   \Big\}   \bigg),
\end{multline}
and the shorthand notation for $*=k,\bar{k}$,
\begin{equation}\label{notationk}
XT_a^{(*)}Y=X_{i_1\dots j \dots i_d}^{j_1\dots j_d}(T_a)_l^{j}Y^{i_1 \dots l\dots i_d}_{j_1\dots j_d}.
\end{equation}
For later convenience, let us define
\begin{eqnarray}\label{Bdefinition}
B^k_a(x)&=&\overline{v}T_a^{(k)}\Phi(x)-\overline{\Phi}(x)T_a^{(k)}v,\nonumber \\
B^{\bar{k}}_a(x)&=&\Phi(x)T_a^{(\bar{k})}\overline{v}-vT_a^{(\bar{k})}\overline{\Phi}(x),\qquad k,\bar{k}=1,\dots,d.
\end{eqnarray}
As we will see later on, the fields $B^k_a(x)$ and $B^{\bar{k}}_a(x)$  are  indeed the Goldstone bosons,  intimately related to the matrix modes. The Goldstone boson definition \eqref{Bdefinition} is a general result, which holds for any Lie group of symmetry. 

When we consider SSB into $G_\omega(N)$ as in \eqref{SSBdiag} there will be $\omega$ constraints. These constraints appear as we perform the transformation of the field according to the SSB pattern $\prod_{\alpha=1}^k\text{Diag}[H_{\alpha}]$, and derive with respect to the parameters. Since the diagonal group $\text{Diag}[H_{\alpha}]$ ``identifies'' the unitary groups within it, there will be only one parameter for each diagonal group. So, for a SSB  pattern  we have a collection of parameters $\{\theta_a^{\alpha}|~\alpha=1,\dots,\omega\}$. The $\omega$ constraints are obtained from
\begin{equation}\label{derivatingZgen}
\frac{\partial Z(J(x),\overline{J}(x))}{\partial \theta^\alpha_a}=0,\qquad \alpha=1,\dots,\omega.
\end{equation}

Now, let us apply the transformations of the diagonal group defined as
\begin{equation}\label{diagaction}
\Phi^{i_1\dots i_d}_{j_1 \dots j_d}(x)\longrightarrow g_{m_1}^{i_1}\cdots g_{m_d}^{i_d} (g^\dagger)_{j_1}^{l_1}\cdots (g^\dagger)_{j_d}^{l_d}\Phi^{m_1 \dots m_d}_{l_1 \dots l_d}(x),
\end{equation}
for elements $g$ of the diagonal action of $U(N)$ on $\Phi$.
The transformation \eqref{diagaction} and the subsequent derivations \eqref{derivatingZ}, but in this case with $\theta_a^{1}=\dots=\theta_a^{d}=\theta_a^{\bar{1}}=\dots =\theta_a^{\bar{d}}=\theta_a$, 
\begin{equation}\label{derivatingZdig}
\frac{\partial Z(J(x),\overline{J}(x))}{\partial \theta_a}=0,
\end{equation}
 lead to the ``diagonal'' constraint 
\bea\label{basicidentitiesdiagonal}
&&\sum_{k=1}^{d}\int d^4x \langle \overline{J}(x)T_a^{(k)}\Phi(x)-\overline{\Phi}(x)T_a^{(k)}J(x)\rangle_{J,\epsilon}-\nonumber\\ 
&&\sum_{\bar{k}=1}^{d}\int d^4x \langle \Phi(x)T_a^{(\bar{k})}\overline{J}(x)-J(x)T_a^{(\bar{k})}\overline{\Phi}(x)\rangle_{J,\epsilon}=0,
\eea
where we have used the property $T_a^\dagger=T_a$. Be aware that no terms depending on $\epsilon$ do appear in the equation \eqref{basicidentitiesdiagonal} in contrast to equations \eqref{basicidentities} and \eqref{basicidentities2}. This happens because we are assuming that for certain shapes of $v$, as discussed in section \ref{SSpatterns}, the term $\epsilon |\Phi(x)-v|^2$ is invariant under the diagonal transformation. The non-appearance  of the $\epsilon$-term is common for other SSB patterns, since it only reflects the fact that the $\epsilon$-term is invariant under the remaining group.

 Using \eqref{basicidentities} and \eqref{basicidentities2} we can rewrite the constraint \eqref{basicidentitiesdiagonal} as
\bea\label{basicidentitiesdiagonal22}
&&\sum_{k=1}^{d}\int d^4x \langle \big(\overline{v}T_a^{(k)}\Phi(x)-\overline{\Phi}(x)T_a^{(k)}v\big)\rangle_{J,\epsilon}-\nonumber\\ 
&&\sum_{\bar{k}=1}^{d}\int d^4x \langle \big(\Phi(x)T_a^{(\bar{k})}\overline{v}-vT_a^{(\bar{k})}\overline{\Phi}(x)\big)\rangle_{J,\epsilon}=0,
\eea
which must be satisfied for all $\Phi$ and $\bar{\Phi}$ and $\forall a$. 
Using the definition of the Goldstone bosons \eqref{Bdefinition}, the constraint \eqref{basicidentitiesdiagonal22} reads 
\be\label{basicidentitiesdiagonal22G.B}
 \Big{\langle}  \int d^4x\Big(\sum_{k=1}^{d}B_a^k(x)-
\sum_{\bar{k}=1}^{d} B_a^{\bar{k}}(x)\Big)\Big{\rangle}_{J,\epsilon}=0.
\ee
Notice that  \eqref{basicidentitiesdiagonal22} imposes strong restrictions on $v$. The invariant tensors $v$ considered in section \ref{SSpatterns} for SSB into the diagonal group fulfill equation   \eqref{basicidentitiesdiagonal22G.B}, which can be taken as a consistency check. In section \ref{(in)dependence} we show, for a generic SSB (i.e., a generic $v$), how constraints of the type \eqref{basicidentitiesdiagonal22G.B} are fulfilled.

\subsection{The Ward-Takahashi identities}
Now, the WT identities, which relate Green functions, are obtained by functional differentiating (\ref{basicidentities}) and \eqref{basicidentities2}  repeatedly with respect to the sources $J(y)$ and $\overline{J}(y)$ and then setting $J(y)=\overline{J}(y)=0$. For example, if we operate with $\frac{\delta}{\delta \bar{J}^{m_1\dots m_d}_{n_1\dots n_d}(y)}\big|_{J(y)=\overline{J}(y)=0}$ on \eqref{basicidentities} and \eqref{basicidentities2}, respectively, we obtain
\begin{eqnarray}\label{1point}
(T_a)^{n_k}_j\langle \Phi^{n_1\dots j\dots n_d}_{m_1\dots m_d}(y)\rangle_\epsilon&=&-\epsilon\int d^4x~\langle \Phi^{n_1\dots n_d}_{m_1\dots m_d}(y),\big(\overline{v}T_a^{(k)}\Phi(x)-\overline{\Phi}(x)T_a^{(k)}v\big)\rangle_{\epsilon},\nonumber \\
(T_a)_{m_k}^j\langle \Phi^{n_1\dots n_d}_{m_1\dots j\dots m_d}(y)\rangle_\epsilon&=&-\epsilon\int d^4x~\langle \Phi^{n_1\dots n_d}_{m_1\dots m_d}(y),\big(\Phi(x)T_a^{(\bar{k})}\overline{v}-vT_a^{(\bar{k})}\overline{\Phi}(x)\big)\rangle_{\epsilon},\nonumber \\
\end{eqnarray}
Here, we have written $\langle F(\Phi)\rangle_\epsilon= \langle F(\Phi)\rangle_{J=0, \epsilon} $. 

Now, following the steps of the previous example of the scalar field, we will take linear combinations of differential operators as in \eqref{deri2}, 
\begin{eqnarray}
\overline{v}T^{(k)}_b\frac{\delta}{\delta \overline{J}(y)}\bigg|_{J(y)=\overline{J}(y)=0}\mp \frac{\delta}{\delta J(y)}\bigg|_{J(y)=\overline{J}(y)=0}T^{(k)}_b v,\label{doperators1}\\
\frac{\delta}{\delta \overline{J}(y)}\bigg|_{J(y)=\overline{J}(y)=0}T^{(\bar{k})}_b\overline{v}\mp vT^{(\bar{k})}_b\frac{\delta}{\delta J(y)}\bigg|_{J(y)=\overline{J}(y)=0}\label{doperators2},
\end{eqnarray} 
onto (\ref{basicidentities}) and (\ref{basicidentities2}) . The subtraction in \eqref{doperators1} applied on (\ref{basicidentities}) results in
\begin{multline}
\langle (T_a^{(k)})^l_{j}(T^{(k)}_b)_l^{i}\Phi^{j_1\dots j\dots j_d}_{i_1\dots i_d}(y)\overline{v}_{j_1\dots i\dots j_d}^{i_1\dots i_d} \\+(T_b^{(k)})^l_{i}(T^{(k)}_a)_l^{j}\overline{\Phi}^{i_1\dots i_d}_{j_1 \dots j\dots j_d}(y)v^{j_1\dots i\dots j_d}_{i_1\dots i_d}\rangle_\epsilon 
=-\epsilon\int d^4x~\langle B_b^k(y), B_a^k(x)\rangle_{\epsilon}.
\end{multline}
We now apply the product rule of generators of $U(N)$
\be\label{prodrule} 
T_aT_b=\frac{1}{2}\text{i}f_{abc}T_c+\frac{1}{2}d_{abc}T_c,\quad;\quad a,b,c=0,1,...,N^2-1,
\ee
where  $f_{abc}$ are the structure constants, and $d_{abc}$ is a totally symmetric tensor.

\begin{equation}\label{WTidentity1}
\frac{1}{2}\text{i}f_{abc}\langle B_c^k(y)\rangle_\epsilon
+\frac{1}{2}d_{abc}\langle \varphi_c^k(y)\rangle_\epsilon
=-\epsilon\int d^4x~\langle B_a^k(y),B_b^k(x) \rangle_{\epsilon},
\end{equation}
where we have used the notation of \eqref{notationk}.

The addition operator in  \eqref{doperators1} applied on (\ref{basicidentities}) leads to the WT identity
\begin{equation}\label{WTidentity2}
\frac{1}{2}\text{i}f_{abc}\langle \varphi_c^k(y)\rangle_\epsilon
+\frac{1}{2}d_{abc}\langle B_c^k(y)\rangle_\epsilon=-\epsilon\int d^4x~\langle \varphi_a^k(y), B_b^k(x)\rangle_{\epsilon},
\end{equation}
where we have defined 
\bea\label{phis}
\varphi_a^k(y)&=&\overline{v}T_a^{(k)}\Phi(x)+\overline{\Phi}(x)T_a^{(k)}v,\nonumber \\
\varphi_a^{\bar{k}}(y)&=&\Phi(x)T_a^{(\bar{k})}\overline{v}+vT_a^{(\bar{k})}\overline{\Phi}(x),\qquad k,\bar{k}=1,\dots,d.
\eea
Identical equations to \eqref{WTidentity1} and \eqref{WTidentity2} hold for  $B_a^{\bar{k}}$ and $\varphi_a^{\bar{k}}$  defined in \eqref{Bdefinition}  and \eqref{phis} after taking the addition and subtraction combinations in \eqref{doperators2}.

\subsection{The Goldstone modes}
We have named the fields $B_a^k(y)$   and $B_a^{\bar{k}}(y)$, defined in \eqref{Bdefinition}, Goldstone bosons. We should check, however,  that they are massless.
We are going to argue that this is the case using similar arguments as in section \ref{scalar0}. 

In momentum space, the propagators that appear on the RHS of \eqref{WTidentity1} are written as
\begin{equation}\label{propagatorsB}
\langle B^k_a(x)  B^k_a(y)\rangle_\epsilon=\text{i}(2\pi)^{-4}\int d^4p \frac{Z_{B^k_a}}{p^2-m_{B^k_a}^2+\text{i}\epsilon a_{B^k_a}}e^{-\text{i}p(x-y)}+\text{(regular contributions)},
\end{equation}
where $Z_{B^k_a}$ and $a_{B^k_a}$ are renormalization constants. Now, because of the term $e^{-\text{i}p(x-y)}$ in \eqref{propagatorsB}, the integral over $x$ appearing in \eqref{WTidentity1} picks a pole at $p=0$. As in the scalar case, the LHS of \eqref{WTidentity1} is non-vanishing\footnote{In analogy with the scalar case, we will take the fields $\varphi_a^k(y)$ to have non-vanishing expectation values. No further assumptions need to be taken with respect to the expectation value of the Goldstone bosons.}. 
So, the limit $\epsilon \to 0$ of the RHS of \eqref{WTidentity1} has to be non-zero. It implies that   $B^k_a(x)$ is massless. That is,
\begin{equation}
 \lim_{\epsilon\to 0}\epsilon\int d^4x ~\langle B^k_a(x)  B^k_a(y) \rangle_\epsilon=-\delta_{ab}\frac{Z_{B^k_a}}{a_{B^k_a}}, \quad m_{B^k_a}=0,
\end{equation}
which is nothing but the Goldstone theorem: when the symmetry gets spontaneously broken, to each broken generator corresponds a massless modes.

\subsection{Linear (in)dependence of the Goldstone modes} \label{(in)dependence}
For a generic SSB pattern $\prod_{\alpha=1}^\omega\text{Diag}[H_{\alpha}]$, we will have $\omega$ constraints given by the $\omega$ equations \eqref{derivatingZgen}. In terms of the Goldstone bosons, those equations are
\be\label{basicidentitiesgeneral22G.BB}
 \Big{\langle}  \int d^4x\Big(\sum_{k'=1}^{n_\alpha}B_a^{\alpha,k'}(x)-
\sum_{\bar{k}'=1}^{n_\alpha} B_a^{\alpha, \bar{k}'}(x)\Big)\Big{\rangle}_{J,\epsilon}=0,\qquad \alpha=1,\dots,\omega,
\ee
where 
\be\label{reorganization}
B_a^{\alpha,k'}(x)\equiv B_a^{k}(x),\qquad B_a^{\alpha,\bar{k}'}(x)\equiv B_a^{\bar{k}}(x),
\ee
is a reorganization of the labels, related to the partition $\sum_{\alpha=1}^{\omega}n_\alpha=d$.
Notice that equation \eqref{basicidentitiesdiagonal22G.B}, for the full diagonal group, is a particular case of \eqref{basicidentitiesgeneral22G.BB} when $\omega=1$. 

We are going to see that \eqref{basicidentitiesgeneral22G.BB} are automatically satisfied provided
 \be\label{basicidentitiesgeneral22G.BB2}
\sum_{k'=1}^{n_\alpha}B_a^{\alpha,k'}(x)-
\sum_{\bar{k}'=1}^{n_\alpha} B_a^{\alpha, \bar{k}'}(x)=0,\qquad \alpha=1,\dots,\omega,
\ee
which, as we are going to show, is a consequence of the definition of the Goldstone bosons \eqref{Bdefinition}.

Each of the equations of \eqref{basicidentitiesgeneral22G.BB2} is associated to a diagonal group $\text{Diag}[H_{\alpha}]$. So the discussion will focus on 
a particular $H_{\alpha}$ (or equivalently, a single cycle in the intersection-diagram of length $n_\alpha$). 

In order not to overload the paper with notation we will consider the case $n_\alpha=3$ for a given $\alpha$.  Without loss of generality the cycle of length 3 in $v$ under consideration  can be chosen 
\be\label{v3cycle}
v^{i_1\dots i_d}_{j_1\dots j_d}=v_1\delta^{i_1i_2i_3\dots i_d}_{j_2j_1j_3\dots j_d}+v_2\delta^{i_1i_2i_3\dots i_d}_{j_2j_1j_3\dots j_d},
\ee
where the indices $4,\dots,d$ correspond to the other diagonal groups in $\prod_{\alpha=1}^\omega\text{Diag}[H_{\alpha}]$, and appear in the rest of the equations of \eqref{basicidentitiesgeneral22G.BB2}. 

According to the definition \eqref{Bdefinition} and the choice \eqref{v3cycle}, the Goldstone bosons are
\begin{eqnarray}\label{rainbow}
B^{\alpha,1}_a(x)&=&{\color{red} v_1(T_a)^l_j\Phi^{jj_2j_3\dots j_d}_{j_2lj_3\dots j_d}(x)}-{\color{blue}v_2(T_a)^l_j\Phi^{jj_2j_3\dots j_d}_{lj_3j_2\dots j_d}(x)}-\text{c.c.},\nonumber \\
B^{\alpha,2}_a(x)&=& {\color{brown}v_1(T_a)^l_j\Phi^{j_1jj_3\dots j_d}_{lj_1j_3\dots j_d}(x)}-{\color{teal}v_2(T_a)^l_j\Phi^{j_1jj_3\dots j_d}_{j_1j_3l\dots j_d}(x)}-\text{c.c.},\nonumber \\
B^{\alpha,3}_a(x)&=& {\color{violet}v_1(T_a)^l_j\Phi^{j_1j_2j\dots j_d}_{j_2j_1l\dots j_d}(x)}-{\color{darkgray}v_2(T_a)^l_j\Phi^{j_1j_2j\dots j_d}_{j_1lj_2\dots j_d}(x)}-\text{c.c.},\nonumber \\
B^{\alpha,\bar{1}}_a(x)&=& {\color{brown}v_1(T_a)^l_j\Phi^{j_2jj_3\dots j_d}_{lj_2j_3\dots j_d}(x)}-{\color{blue}v_2(T_a)^l_j\Phi^{jj_3j_2\dots j_d}_{lj_2j_3\dots j_d}(x)}-\text{c.c.},\nonumber \\
B^{\alpha,\bar{2}}_a(x)&=& {\color{red}v_1(T_a)^l_j\Phi^{jj_1j_3\dots j_d}_{j_1lj_3\dots j_d}(x)}-{\color{darkgray}v_2(T_a)^l_j\Phi^{j_1j_3j\dots j_d}_{j_1lj_3\dots j_d}(x)}-\text{c.c.},\nonumber \\
B^{\alpha,\bar{3}}_a(x)&=& {\color{violet}v_1(T_a)^l_j\Phi^{j_2j_1j\dots j_d}_{j_1j_2l\dots j_d}(x)}-{\color{teal}v_2(T_a)^l_j\Phi^{j_1jj_2\dots j_d}_{j_1j_2l\dots j_d}(x)}-\text{c.c.}\ .
\end{eqnarray}
The constraint equation for the group $\text{Diag}[H_{\alpha}]$ is 
\begin{equation}\label{cancellation}
B^{\alpha,1}_a(x)+B^{\alpha,2}_a(x)+B^{\alpha,3}_a(x)-B^{\alpha,\bar{1}}_a(x)-B^{\alpha, \bar{2}}_a(x)-B^{\alpha,\bar{3}}_a(x)=0.
\ee
This equation is automatically fulfilled from \eqref{rainbow}. The cancellations among the Goldstone bosons have been depicted in colors. The crucial point is that the cancellation \eqref{cancellation} occurs only when {\it all} the Goldstone bosons are involved as we can see in \eqref{cancellation}. No partial cancellations like $B^{\alpha,1}_a(x) -B^{\alpha,\bar{3}}_a(x)=0$ or $B^{\alpha,1}_a(x)+B^{\alpha,2}_a(x)-B^{\alpha,\bar{1}}_a(x)-B^{\alpha,\bar{2}}_a(x)=0$ happen, which implies that only one Goldstone boson in \eqref{rainbow} is linearly dependent. This is general feature for any of the $\omega$ groups $\text{Diag}[H_{\alpha}]$: for each $\text{Diag}[H_{\alpha}]$ there are $2n_\alpha-1$ linearly independent Goldstone bosons $B_a(x)$, where $a=0,\dots, N^2-1$.

As a remark, there is a relation between the cancellation pattern shown in colors in   \eqref{rainbow} and the  cycle corresponding to $H_{\alpha}$ in the intersection-diagram. Note that drawing lines that join the same colors, the whole set of equations \eqref{rainbow} form one and only one loop.  The similarity between these pictures goes beyond this observation, and suggests a deep connection which will be studied elsewhere. 

Let us count the total number of Goldstone bosons in a SSB pattern $\prod_{\alpha=1}^\omega\text{Diag}[H_{\alpha}]$. As said above, for each $\text{Diag}[H_{\alpha}]$ there are $(2n_\alpha-1)N^2$ Goldstone bosons. So for $\prod_{\alpha=1}^\omega\text{Diag}[H_{\alpha}]$ there are
\be \label{countingcool}
\sum_{\alpha=1}^\omega (2n_\alpha-1)N^2=(2d-\omega)N^2,
\ee
Goldstone bosons, where we have used $\sum_{\alpha=1}^\omega n_\alpha=d$. Equation\footnote{There is a remarkable matching between the counting in equation \eqref{countingcool} for the complete breaking of the symmetry ($\omega=0$)  and the number of light modes reported in \cite{Choudhury:2017tax} for $O(N)$. In the case of $O(N)$ with $d$ even, the total number of indices of the tensor $\phi_{i_1\dots i_d}$ is $d$, instead of $2d$ in equation \eqref{countingcool}. Moreover, since the tensor field for the $O(N)$ group is real, the number of modes is half the number of the unitary case. With these considerations, the counting of the Goldstone modes for the orthogonal group, when the SSB breaks completely the original symmetry, is $\frac{1}{2}dN^2$, which is the number of light modes counted in p.5 of \cite{Choudhury:2017tax}, with the identification $d=q-1$. }  \eqref{countingcool} is in perfect agreement with the direct counting of the number of broken symmetries in equation \eqref{conteo}.  

\subsection{Matrix organization of the Goldstone modes}\label{matrixGB}
So far, we have described the appearance of the Goldstone bosons $B_a^{\alpha,k'}(x)$ and $B^{\alpha,\bar{k}'}_a(y)$ as a consequence of the SSB $G_{d\bar{d}}(N)\to \prod_{\alpha=1}^\omega\text{Diag}[H_{\alpha}]$. The number of Goldstone modes match the number of broken continuous symmetries, which is $(2d-\omega)N^2$. So in the effective theory we have a collection of linearly independent fields
\begin{equation}
B_\alpha=\{B_a^{\alpha,k'}(x),B_a^{\alpha,\bar{k}'}(x)\big|~a=0,\dots,N^2-1;~~k'=1,\dots,n_\alpha,~~\bar{k}'=1,\dots,n_\alpha-1\},
\end{equation}
where $\alpha=1,\dots,\omega$, and we have chosen the last fields  $B_a^{\alpha,\bar{n}_\alpha}(x)$ to be linearly dependent on the others, according to the constraints \eqref{basicidentitiesgeneral22G.BB2}.

The question now is how these modes organize into multiplets. To answer this question we will consider an action of the group $\text{Diag}[H_\alpha]$ on the space  $B_\alpha$, treated as a vector space, and find the irreducible representations of $B_\alpha$. The Goldstone modes multiplets will correspond, one-to-one, to those irreps.  

The action of $\prod_{\alpha=1}^\omega\text{Diag}[H_{\alpha}]$ on the tensor field induces the transformation on each $B_\alpha$ 
\begin{eqnarray}\label{gactiononB}
B_a^{\prime\alpha, k'}(x)&=&\overline{v} [g_\alpha T_a^{(k)}g_\alpha^\dagger]\Phi(x)-\overline{\Phi}(x)[g_\alpha T_a^{(k)}g_\alpha^\dagger] v, \nonumber \\
B_a^{\prime\alpha, \bar{k}'}(x)&=&\Phi(x) [g_\alpha T_a^{(\bar{k})}g_\alpha^\dagger]\overline{v}-v[g_\alpha T_a^{ (\bar{k})}g_\alpha^\dagger] \overline{\Phi}(x),\qquad g_\alpha\in \text{Diag}[H_\alpha],
\end{eqnarray}
where the labels $(\alpha,k')$ and $k$ are related according to the map \eqref{reorganization}.
Notice that the action of $g_\alpha$ is trivial in all the slots except for the slot $k$, where the generators  hit.
So the action of $\text{Diag}[H_\alpha]$ on $B_\alpha$ reduces to the adjoint action 
\begin{equation}
\text{Ad}_{g_\alpha} T_a=g_{\alpha}T_ag_{\alpha}^\dagger=\sum_bC^b_a(g_{\alpha}) T_b,\quad g_{\alpha}\in \text{Diag}[H_\alpha],\quad T_a\in   \mathfrak{u}(N),
\end{equation}
on each of the  slots labeled by $k$ and $\bar{k}$. \\

So for each $(\alpha,k')$ and $(\alpha,\bar{k}')$, all the modes $B_a^{\alpha,k'}(x)$ and  $B_a^{\alpha,\bar{k}'}(x)$ with $a=0,\dots,N^2-1$ get arranged into a multiplet. The transformation \eqref{gactiononB} suggests that  $B_a^{\alpha,k'}(x)$ and  $B_a^{\alpha,\bar{k}'}(x)$ are the components of an $N\times N$-matrix field. In fact, we can map each $(\alpha, k')$ and $(\alpha, \bar{k}')$ collection $\{B_a^{\alpha, k'}(x),B_a^{\alpha,\bar{k}'}(x)|~a=0,\dots,N^2-1\}$ into the matrices 
\begin{equation}\label{Zdefined}
(B^{\alpha,k'})^i_j(x)= \sum_a B_a^{\alpha,k'}(x)(T_a)^i_j,\qquad k'=1,\dots, n_\alpha,
\end{equation} and 
\begin{equation}\label{Zdefined2}
(B^{\alpha,\bar{k}'})^i_j(x)= \sum_a B_a^{\alpha, \bar{k}'}(x)(T_a)^i_j,\qquad \bar{k}'=1,\dots, n_\alpha-1.
\end{equation}
These fields transform in the adjoint of $U_\alpha(N)=\text{Diag}[H_\alpha]$ as 
\begin{multline}\label{Ztransform}
(B'^{\alpha,k'})^i_j(x)=(g_\alpha)^i_m(B^{\alpha,k'})^m_l(x)(g_\alpha^\dagger)^l_j=\\
\sum_{a,b} C^b_a(g_\alpha)B_a^{\alpha,k'}(x)(T_b)^i_j=\sum_b B'^{\alpha,k'}_b(x)(T_b)^i_j,
\end{multline}
where 
\begin{equation}\label{B'}
B'^{\alpha,k'}_b(x)=\sum_a C^a_b(g_{\alpha})B^{\alpha,k'}_a(x).
\end{equation}
The same transformation law holds for $(\alpha,\bar{k}')$ fields.

The transformation \eqref{Ztransform} is perfectly compatible with \eqref{gactiononB}. Indeed,
\begin{eqnarray}\label{gactiononBcompatible}
 B_b^{\prime \alpha,k'}(x)&=&\overline{v} [g_{\alpha} T_b^{(k)}g_{\alpha}^\dagger]\Phi(x)-\overline{\Phi}(x)[g_{\alpha}T_b^{(k)}g_{\alpha}^\dagger] v\nonumber \\
&=&\sum_aC^a_b(g_{\alpha})\big(\overline{v} T_a^{(k)}\Phi(x)-\overline{\Phi}(x)T_a^{(k)} v\big)\nonumber \\
&=&\sum_aC^a_b(g_{\alpha}) B^{\alpha,k'}_a(x),
\end{eqnarray}
which is the transformation law \eqref{B'}. We have shown that the Goldstone bosons $B^{\alpha,k'}_a(x)$ and $B^{\alpha,\bar{k}'}_a(x)$, with $a=0,\dots,N^2-1,$ are actually the components of a matrix transforming in the adjoint of $U_\alpha(N)$.

Interestingly enough, we can write down the Goldstone boson matrix fields  in terms the tensor fields and $v$ exclusively, without the use of the generators\footnote{Although equation \eqref{genmatrix} seems to involve the generators of the group, they appear in the combination $T_aT_a$. Using the product rule for the given group, the explicit form of the generators is not needed, see equation \eqref{mainresult}.}. Using the definitions \eqref{Zdefined} and \eqref{Bdefinition}, we have
\begin{mdframed}
\begin{eqnarray}\label{genmatrix}
(B^{\alpha,k'})^i_j(x)&=& \sum_a B_a^{\alpha,k'}(x)(T_a)^i_j=\sum_a\big(\overline{v} T_a^{(k)}\Phi(x)-\overline{\Phi}(x)T_a^{(k)} v\big)(T_a)^i_j ,\nonumber \\
(B^{\alpha,\bar{k}'})^i_j(x)&=& \sum_a B_a^{\alpha,\bar{k}'}(x)(T_a)^i_j=\sum_a\big(\Phi(x)T_a^{(\bar{k})}\overline{v}-vT_a^{(\bar{k})}\overline{\Phi}(x)\big)(T_a)^i_j.
\end{eqnarray}
\end{mdframed}
{\it Equation \eqref{genmatrix} is completely general, valid for any symmetry group and any SSB. It means that the collection of Goldstone bosons which result from SSB of tensor theories  organizes into matrix field multiplets. In other words, any tensor theory with SSB always leads to matrix theories.} See appendix  \ref{tetraap} for an application of \eqref{genmatrix} to a non-unitary group and a different SSB pattern.

Using the properties of the generators of $U(N)$
\be\label{prodgen}
(T_a)^m_n(T_a)^k_i=\frac{1}{2}\delta_i^m\delta_n^k,
\ee
we arrive at
\begin{mdframed}
\begin{eqnarray}\label{mainresult}
(B^{\alpha,k'})^i_j(x)&=&\frac{1}{2}\Big(\overline{v}^{i_1\dots i_d}_{j_1\dots j\dots j_d}\Phi^{j_1 \dots i \dots j_d}_{i_1\dots i_d}(x)-v^{j_1\dots i\dots j_d}_{i_1\dots i_d}\overline{\Phi}^{i_1\dots  i_d}_{j_1\dots j \dots j_d}(x)\Big), \nonumber \\
(B^{\alpha,\bar{k}'})^i_j(x)&=&\frac{1}{2}\Big(\Phi^{i_1\dots i_d}_{j_1\dots j\dots j_d}\overline{v}^{j_1 \dots i \dots j_d}_{i_1\dots i_d}(x)-\overline{\Phi}^{j_1\dots i\dots j_d}_{i_1\dots i_d}v^{i_1\dots  i_d}_{j_1\dots j \dots j_d}(x)\Big), 
\end{eqnarray}
\end{mdframed}
where it is understood that the indices $i,j$ are located at the $k,\bar{k}$ slots of the tensor. Recall that $k,\bar{k}$ are given in terms of $(\alpha,k')$ and $(\alpha,\bar{k}')$ by the map \eqref{reorganization}. Equations \eqref{mainresult} are one of the main results of the paper. We would like to stress that the Goldstone bosons \eqref{mainresult} are defined  for {\it any } SSB pattern induced by $v$ in unitary groups, where $v$ is not necessarily constrained to the shape considered in this paper, namely linear combinations of Kronecker deltas. 

The fields $\varphi^k_a(x)$ and $\varphi^{\bar{k}}_a(x)$ defined in \eqref{phis} organise into matrix multiplets, for the same reason as the Goldstone bosons.  Although these fields will be massive in the effective theory, hence they will not be considered at low energies.

At low energies, the massless modes (which we assume to be only the Goldstone bosons \cite{Weinberg:1996kr}) are the most relevant. According to the above discussion:\\
{\it The massless field content of the effective theory that results from SSB of a tensor theory built on $\Phi^{i_1\dots i_d}_{j_1\dots j_d}(x)$, into $\prod_{\alpha=1}^\omega\text{Diag}[H_{\alpha}]$, which is associated with the partition $\sum_{\alpha=1}^\omega n_\alpha=d$, is an overall of  $2d-\omega$ $N\times N$-matrix fields organized as $B^{\alpha,k'}(x)$ and $B^{\alpha,\bar{k}'}(x)$, with $k'=1,\dots,n_\alpha$ and $\bar{k}'=1,\dots,n_\alpha-1$, transforming in the adjoint of $U_\alpha(N)=\text{Diag}[H_\alpha]$, and $\alpha=1,\dots,\omega$.}

The fact that the effective theory is invariant under $\prod_{\alpha=1}^\omega U_\alpha(N)$ tells us that the fields $B^{\alpha,k'}(x)$ and $B^{\alpha,\bar{k}'}(x)$ must appear in the action of the effective theory with all the indices properly contracted. So, the vertices of the theory must appear as multi trace polynomials, where the matrix fields appearing within a given trace must involve only fields transforming under the same group, that is, with the same $\alpha$. In addition, the fact that the fields are complex tells us that the monomials that appear in the action must include $B^{\alpha,k'}(x)$ and $\overline{B}^{\alpha,k'}(x)$ in an equal number. So, the action of the effective theory will contain interaction monomials of the type
\begin{equation}
\text{Tr}(B^{1,1}\overline{B}^{1,2})\text{Tr}(\overline{B}^{3,1}B^{3,2})\text{Tr}(\overline{B}^{2,1}B^{2,2}\overline{B}^{2,3}B^{2,4})\cdots ,
\end{equation}
where, for simplicity, we are omitting the spacetime derivatives of the matrix fields.

 This concludes our discussion about the multiplets in the effective field theory. Apart form these particular considerations for our case, the general features of effective field theories \cite{Weinberg:1980wa,DHoker:1994rdl,Weinberg:1997rv} apply as well.

\section{\label{sec:2}Examples}
In this section we are going to apply the general formalism of SSB in tensor theories developed above in two examples. 

\subsection{SSB for the complex 2-tensor field}\label{E2}

In order to make  the discussion of the previous sections more explicit, using the techniques presented above, we shall study the well-known example of chiral symmetry breaking for the complex scalar field, see, for instance, \cite{Bai:2017zhj}. This involves SSB of a complex 2-tensor field $\Phi^{i}_{j}(x)$ transforming in the fundamental and anti-fundamental  representation  $G_{1\bar{1}}(N)= U_1(N)\times U_{\bar{1}}(N)$ into the diagonal group. We will use the notation defined in \eqref{2ind}.\\
The elements of the group  may be written as $(g_1,g_{\bar{1}})$, with
\begin{eqnarray}
g_i&=&\text{exp}(\text{i}\theta^{(i)}_aT_a)\in U_i(N),\nonumber \\
 g^{\dagger}_i&=&\text{exp}(-\text{i}\theta^{(i)}_aT^{\dagger}_a)\in U_i(N),\quad i=1,\bar{1}. \label{nonabeliantrans}
\end{eqnarray}
The generators $T_a$ of the unitary group are Hermitian, so $T_a=T^{\dagger}_a$. However, we will not make the substitution at this stage in order to keep track of the conjugate terms. 
For the 2-tensor case we will require an action and a measure  invariant under the  transformations \eqref{2ind}.
The generating functional $Z_{\epsilon}[J,\overline{J}]$ is defined in the same fashion as for the scalar case in section \ref{scalar0}, 
\begin{multline}
Z_{\epsilon}[J,\overline{J}]=\frac{1}{N}\int D\Phi ~\text{exp}\Big[\text{i}S\big[\Phi(x)\big]+\text{i}\int d^4 x\big( \overline{J}^{i}_{j}(x)\Phi^{j}_{i}(x)+c.c.\big)\\
-\epsilon\int d^4x\big((\Phi(x)-v)^{i}_{j}(\overline{\Phi}(x)-\overline{v})^{j}_{i}\big) \Big]. \label{WepsilonU}
\end{multline}
Here we are only interested in the breaking of the symmetry to the diagonal group, which occurs when  $v^{i}_{j}$ is proportional to $\delta^{i}_{j}$, as discussed in section \ref{SSpatterns}. 

 Following the methodology of previous sections, we will compute the WT identities. 
First, we transform the fields as in  \eqref{2ind} with the elements of the groups written as \eqref{nonabeliantrans}, and derive \eqref{WepsilonU} with respect to the parameters $\theta^{(i)}_a$. The path integral is invariant under the field transformation, so
\be
\frac{\partial}{\partial \theta^{(i)}_a}Z_{\epsilon}[J,\overline{J}]=0,\quad i=1,\bar{1}.
\ee
For $\theta^{(1)}_a$ we obtain
\begin{eqnarray}\label{firstid}
&&\int d^4x \langle \overline{J}^{i}_{j}(T^{\dagger}_a)^{j}_{k}\Phi^{k}_{i}(x)-J^{i}_{j}(T_a)^{k}_{i}\overline{\Phi}^{j}_{k}(x)\rangle_{\epsilon}\nonumber \\
&&-\text{i}\epsilon\int d^4x \langle\overline{v}^{i}_{j}(T^{\dagger}_a)^{j}_{k}\Phi^{k}_{i}(x)-v^{i}_{j}(T_a)^{k}_{i}\overline{\Phi}^{j}_{k}(x)\rangle_{\epsilon}=0, 
\end{eqnarray}
and for $\theta^{(\bar{1})}_a$ we obtain the identity
\begin{eqnarray}\label{secondid}
&&\int d^4x \langle \overline{J}^{i}_{j}(T_a)^{k}_{i}\Phi^{j}_{k}(x)-J^{i}_{j}(T^{\dagger}_a)^{j}_{k}\overline{\Phi}^{k}_{i}(x)\rangle_{\epsilon}\nonumber \\
&&-\text{i}\epsilon\int d^4x \langle\overline{v}^{i}_{j}(T_a)^{k}_{i}\Phi^{j}_{k}(x)-v^{i}_{j}(T^{\dagger}_a)^{j}_{k}\overline{\Phi}^{k}_{i}(x)\rangle_{\epsilon}=0.
\end{eqnarray}
Again, the identities are obtained by derivating with respect to the sources. So, applying $\frac{\delta}{\delta \overline{J}^m_{n}(y)}\Big|_{J=\overline{J}=0}$ on \eqref{firstid} and on \eqref{secondid} we have
\begin{eqnarray}
0&=&(T_a^\dagger)^n_i\langle \Phi_m^i(y)\rangle_\epsilon+\epsilon\int d^4 x\langle\Phi_m^n(y),\overline{v}^i_j(T^\dagger_a)^j_k\Phi_i^k(x)-v_j^i(T_a)_i^k\overline{\Phi}^j_k(x)\rangle_\epsilon \label{theta1},\\
0&=&(T_a)_m^i\langle \Phi_i^n(y)\rangle_\epsilon+\epsilon\int d^4 x\langle\Phi_m^n(y),\overline{v}^i_j(T_a)^k_i\Phi_k^j(x)-v_j^i(T^\dagger_a)_k^j\overline{\Phi}^k_i(x)\rangle_\epsilon.\label{theta2}
\end{eqnarray}

To obtain the  additional  constraint we derive the path integral \eqref{WepsilonU} with respect to $\theta_a=\theta_a^{(1)}=\theta_a^{(\bar{1})}$, which parametrizes the diagonal group. Notice that under the action of the diagonal group the $\epsilon$-term is invariant. The constraint reads
\bea\label{diagonalconstraint41}
&&~~~\int d^4x \langle\overline{v}^{i}_{j}(T^{\dagger}_a)^{j}_{k}\Phi^{k}_{i}(x)-v^{i}_{j}(T_a)^{k}_{i}\overline{\Phi}^{j}_{k}(x)\rangle_{\epsilon}\nonumber\\
&&-\int d^4x \langle\overline{v}^{i}_{j}(T_a)^{k}_{i}\Phi^{j}_{k}(x)-v^{i}_{j}(T^{\dagger}_a)^{j}_{k}\overline{\Phi}^{k}_{i}(x)\rangle_{\epsilon}=0.
\eea
This holds for any configuration of the tensor field, i.e.,
\be\label{diagonalconstraint42}
v^{j}_{k}(T_a)^i_{j}-v^{i}_{j}(T_a)^{j}_k
=0,
\ee
which is satisfied for  $v^i_j=v \delta^i_j$. In terms of the Goldstone bosons \eqref{Bdefinition}, the constraint \eqref{diagonalconstraint41} and/or \eqref{diagonalconstraint42} follows from the cancellation \eqref{basicidentitiesgeneral22G.BB2}, which in this case is
\be\label{Bes}
B_a^1(x)-B_a^{\bar{1}}(x)=0.
\ee
 This proves the consistency of the choice of $v$ for the case of two indices tensor. This result, namely that $v$ is proportional to the identity,  is well-known in the context of chirality breaking. Nevertheless, the generalization of it that we perform in section  \ref{SSpatterns} has not been reported in the literature, as far as we know.\\

Now, we will use \eqref{theta2} to find an expression for the Goldstone bosons in the case of SSB into the diagonal group. We take $v^i_j=v\delta^i_j$ where, without loss of generality, $v$ is real. We will also use the Hermiticity of the generators $T_a=T_a^\dagger$. So, we have 
\be
(T_a)^i_m\langle \Phi_i^n(y)\rangle_\epsilon=-\epsilon  \int d^4 x\langle \Phi_m^n(y),v\Big((T_a)^k_i\Phi^i_k(x)-(T_a)^k_i\overline{\Phi}^i_k(x)\Big)\rangle_\epsilon.  \label{Gb1}
\ee
Conjugating the above equation we get
\be
(T_a)^m_i\langle \overline{\Phi}_n^i(y)\rangle_\epsilon=\epsilon  \int d^4 x\langle \overline{\Phi}_n^m(y),v\Big((T_a)^k_i\Phi^i_k(x)-(T_a)^k_i\overline{\Phi}^i_k(x)\Big)\rangle_\epsilon.  \label{Gb2}
\ee
We multiply \eqref{Gb1} by $(T_b)^m_n$ and \eqref{Gb2} by $(T_b)^n_m$ to obtain the couple of equations
\begin{eqnarray}
(T_b)^m_n(T_a)^i_m\langle \Phi_i^n(y)\rangle_\epsilon&=&-\epsilon  \int d^4 x\langle (T_b)^m_n\Phi_m^n(y),v(T_a)^k_i\Big(\Phi^i_k(x)-\overline{\Phi}^i_k(x)\Big)\rangle_\epsilon, \label{Gb3} \\ 
(T_b)^n_m(T_a)^m_i\langle \overline{\Phi}_n^i(y)\rangle_\epsilon&=&\epsilon  \int d^4 x\langle (T_b)^n_m\overline{\Phi}_n^m(y),v(T_a)^k_i\Big(\Phi^i_k(x)-\overline{\Phi}^i_k(x)\Big)\rangle_\epsilon. \label{Gb4} 
\end{eqnarray}
Multiplying by $v$ \eqref{Gb3} and \eqref{Gb4}, and summing them  we obtain
\begin{eqnarray}
&&v\Big((T_b)^m_n(T_a)^i_m\langle \Phi_i^n(y)\rangle_\epsilon+(T_b)^n_m(T_a)^m_i\langle \overline{\Phi}_n^i(y)\rangle_\epsilon\Big) \nonumber \\
&&=\epsilon  \int d^4 x\langle v\Big((T_b)^m_n\Phi_m^n(y)-(T_b)^n_m\overline{\Phi}_n^m(y)\Big),v(T_a)^k_i\Big(\Phi^i_k(x)-\overline{\Phi}^i_k(x)\Big)\rangle_\epsilon. \label{Gb5}
\end{eqnarray}
We will  use the multiplication rule for the generators \eqref{prodrule}.
Now we define the fields
\begin{eqnarray}
B_a(x)&=& v(T_a)^k_i\Big(\Phi^i_k(x)-\overline{\Phi}^i_k(x)\Big), \label{Gbfield} \\
\varphi_a(x)&=&v(T_a)^k_i\Big(\Phi^i_k(x)+\overline{\Phi}^i_k(x)\Big). \label{varphifield} 
\end{eqnarray}
Note that the field $\varphi_a(x)$ is real whereas $B_a(x)$ is purely imaginary. 
With these definitions, \eqref{Gb5} can be written as 
\be \label{idGB}
\frac{1}{2}\text{i}f_{abc}\langle B_c(y)\rangle_\epsilon+\frac{1}{2}d_{abc}\langle  \varphi_c(y)\rangle_\epsilon=-\epsilon \int d^4 x\langle B_a(y),B_b(x)\rangle_\epsilon.
\ee
Now, equations \eqref{Gb3} and \eqref{Gb4} could also be subtracted. In that case, and using the same fields as before we obtain the WT identity
\be \label{idvarphi}
\frac{1}{2}\text{i}f_{abc}\langle \varphi_c(y)\rangle_\epsilon+\frac{1}{2}d_{abc}\langle  B_c(y)\rangle_\epsilon=-\epsilon \int d^4 x\langle \varphi_a(y),B_b(x)\rangle_\epsilon.
\ee
In order to identify the effective degree of freedom it is convenient to rewrite (\ref{Gbfield}) and  (\ref{varphifield}) as matrices. To this end we define the new fields 
\begin{eqnarray}
B^m_n(x)&=& B_a(x)(T_a)^m_n, \label{Gbfieldmatrix} \\
\varphi^m_n(x)&=&\varphi_a(x)(T_a)^m_n, \label{varphifieldmatrix} 
\end{eqnarray}
and using the identity 
\be
(T_a)^m_n(T_a)^k_i=\frac{1}{2}\delta_i^m\delta_n^k,
\ee
we get
\begin{eqnarray}
B^m_n(x)&=& \frac{v}{2}\Big(\Phi^m_n(x)-\overline{\Phi}^m_n(x)\Big), \label{Gbfieldm1} \\
\varphi^m_n(x)&=&\frac{v}{2}\Big(\Phi^m_n(x)+\overline{\Phi}^m_n(x)\Big),\label{varphifieldm2} 
\end{eqnarray} 
where now (\ref{Gbfieldm1}) and (\ref{varphifieldm2}) are   anti-hermitian and hermitian  tensors, respectively. As discussed in section \ref{matrixGB},
 the matrix field $B^m_n(x)$ is the Goldstone boson matrix field. Particularly,  \eqref{Gbfieldm1}  is in accordance with \eqref{mainresult}.

In the next section we will study a more involved case where the tensor field has four indices.

\subsection{SSB  for the complex 4-tensor field} \label{E4}

Following the steps of the previous section, we will show how the SSB of $G_{2\bar{2}}(N)= U_1(N)\times U_2(N)\times U_{\bar{1}}(N)\times U_{\bar{2}}(N)$ into the diagonal subgroup is triggered by the $\epsilon$-term. We will also study the Goldstone bosons, which display a much richer structure than the two-tensor field. Using the notation \eqref{unitaryaction} and \eqref{unitaryactionconjugate}, the fields will transform as
\begin{eqnarray} 
\Phi'^{i_1 i_2}_{j_1 j_2}&=&(g_1^\dagger)^{i_1}_{l_1}(g_2^\dagger)^{i_2}_{l_2}(g_{\bar{1}})^{k_1}_{j_1} (g_{\bar{2}})^{k_2}_{j_2}\Phi^{l_1 l_2}_{k_1 k_2},\nonumber \\
\overline{\Phi}'^{i_1 i_2}_{j_1 j_2}&=&(g_{\bar{1}}^\dagger)^{i_1}_{l_1} (g_{\bar{2}}^\dagger)^{i_2}_{l_2}(g_1)^{k_1}_{j_1} (g_2)^{k_2}_{j_2}\overline{\Phi}^{l_1 l_2}_{k_1 k_2},
\end{eqnarray}
where it is manifest that $\Phi'^{i_1 i_2}_{j_1 j_2}\overline{\Phi}'^{j_1 j_2}_{i_1 i_2}$ is invariant under $G_{2\bar{2}}(N)$.\\
The elements of the group may be written as
\begin{eqnarray}
g_i&=&\text{exp}(\text{i}\theta^{(i)}_aT_a)\in U_i(N),\nonumber \\
 g^{\dagger}_i&=&\text{exp}(-\text{i}\theta^{(i)}_aT^{\dagger}_a)\in U_i(N),\quad i=1,2,\bar{1},\bar{2}. \label{nonabeliantrans4}
\end{eqnarray}
The path integral in this case reads
\begin{multline}
Z_{\epsilon}[J,\overline{J}]=\frac{1}{N}\int D\Phi ~\text{exp}\Big[\text{i}S\big[\Phi(x)\big]+\text{i}\int d^4 x\big( \overline{J}^{i_1i_2}_{j_1j_2}(x)\Phi^{j_1j_2}_{i_1i_2}(x)+c.c\big)\\
-\epsilon\int d^4x\big((\Phi(x)-v)^{i_1i_2}_{j_1j_2}(\overline{\Phi}(x)-\overline{v})^{j_1j_2}_{i_1i_2}\big) \Big]. \label{WepsilonU4index}
\end{multline}
We will again transform the fields with the parametrizations \eqref{nonabeliantrans4} and find the four basic identities through the derivatives
\be
\frac{\partial}{\partial\theta^{(i)}_a}Z_{\epsilon}[J,\overline{J}]=0,\quad i=1,2,\bar{1},\bar{2}.
\ee

For $\theta^{(1)}_a$ we obtain
\begin{eqnarray}\label{firstid4}
&&\int d^4x \langle \overline{J}^{i_1i_2}_{jj_2}(T^{\dagger}_a)^{j}_{j_1}\Phi^{j_1j_2}_{i_1i_2}(x)-J^{ii_2}_{j_1j_2}(T_a)^{i_1}_{i}\overline{\Phi}^{j_1j_2}_{i_1i_2}(x)\rangle_{\epsilon}\nonumber \\
&&-\text{i}\epsilon\int d^4x \langle\overline{v}^{i_1i_2}_{jj_2}(T^{\dagger}_a)^{j}_{j_1}\Phi^{j_1j_2}_{i_1i_2}(x)-v^{ii_2}_{j_1j_2}(T_a)^{i_1}_{i}\overline{\Phi}^{j_1j_2}_{i_1i_2}(x)\rangle_{\epsilon}=0, 
\end{eqnarray}
and for $\theta^{(\bar{1})}_a$ we obtain the identity
\begin{eqnarray}\label{secondid4}
&&\int d^4x \langle \overline{J}^{ii_2}_{j_1j_2}(T_a)^{i_1}_{i}\Phi^{j_1j_2}_{i_1i_2}(x)-J^{i_1i_2}_{jj_2}(T^\dagger_a)^{j}_{j_1}\overline{\Phi}^{j_1j_2}_{i_1i_2}(x)\rangle_{\epsilon}\nonumber \\
&&-\text{i}\epsilon\int d^4x \langle\overline{v}^{ii_2}_{j_1j_2}(T_a)^{i_1}_{i}\Phi^{j_1j_2}_{i_1i_2}(x)-v^{i_1i_2}_{jj_2}(T^\dagger_a)^{j}_{j_1}\overline{\Phi}^{j_1j_2}_{i_1i_2}(x)\rangle_{\epsilon}=0, 
\end{eqnarray}
For $\theta^{(2)}_a$, the generator hits on the second position as indicated by $k=2$.  We obtain
\begin{eqnarray}\label{firstid44}
&&\int d^4x \langle \overline{J}^{i_1i_2}_{j_1j}(T^{\dagger}_a)^{j}_{j_2}\Phi^{j_1j_2}_{i_1i_2}(x)-J^{i_1i}_{j_1j_2}(T_a)^{i_2}_{i}\overline{\Phi}^{j_1j_2}_{i_1i_2}(x)\rangle_{\epsilon}\nonumber \\
&&-\text{i}\epsilon\int d^4x \langle\overline{v}^{i_1i_2}_{j_1j}(T^{\dagger}_a)^{j}_{j_2}\Phi^{j_1j_2}_{i_1i_2}(x)-v^{i_1i}_{j_1j_2}(T_a)^{i_2}_{i}\overline{\Phi}^{j_1j_2}_{i_1i_2}(x)\rangle_{\epsilon}=0, 
\end{eqnarray}
and for $\theta^{(\bar{2})}_a$ we obtain the identity
\begin{eqnarray}\label{secondid44}
&&\int d^4x \langle \overline{J}^{i_1i}_{j_1j_2}(T_a)^{i_2}_{i}\Phi^{j_1j_2}_{i_1i_2}(x)-J^{i_1i_2}_{j_1j}(T^\dagger_a)^{j}_{j_2}\overline{\Phi}^{j_1j_2}_{i_1i_2}(x)\rangle_{\epsilon}\nonumber \\
&&-\text{i}\epsilon\int d^4x \langle\overline{v}^{i_1i}_{j_1j_2}(T_a)^{i_2}_{i}\Phi^{j_1j_2}_{i_1i_2}(x)-v^{i_1i_2}_{j_1j}(T^\dagger_a)^{j}_{j_2}\overline{\Phi}^{j_1j_2}_{i_1i_2}(x)\rangle_{\epsilon}=0.
\end{eqnarray}
Note that equation \eqref{firstid44} is analogous to \eqref{firstid4} except for the generator hitting on the second slot of the fields. The same happens to equations \eqref{secondid44} and \eqref{secondid4}.
For a tensor of $2d$ indices we will have $d$ similar equations to \eqref{firstid4} and other $d$ equations similar to \eqref{secondid4}, where the generators hit on each of the $d$ slots.\\

In order to obtain WT identities we derive with respect the sources. So, applying $\frac{\delta}{\delta \overline{J}^{m_1m_2}_{n_1n_2}(y)}\Big|_{\overline{J}=J=0}$ on \eqref{firstid4},  \eqref{secondid4}, \eqref{firstid44} and \eqref{secondid44}   we obtain
\begin{eqnarray}
(T_a^\dagger)^{n_1}_{j}\langle \Phi^{jn_2}_{m_1m_2}(y)\rangle_\epsilon&=&-\epsilon\int d^4 x\langle\Phi_{m_1m_2}^{n_1n_2}(y),\overline{v}^{i_1i_2}_{jj_2}(T^\dagger_a)^j_{j_1}\Phi^{j_1j_2}_{i_1i_2}(x)-v_{j_1j_2}^{ii_2}(T_a)_i^{i_1}\overline{\Phi}^{j_1j_2}_{i_1i_2}(x)\rangle_\epsilon,  \nonumber \\
(T_a)_{m_1}^{i}\langle \Phi_{im_2}^{n_1n_2}(y)\rangle_\epsilon&=&-\epsilon\int d^4 x\langle\Phi_{m_1m_2}^{n_1n_2}(y),\overline{v}^{ii_2}_{j_1j_2}(T_a)^{i_1}_i\Phi_{i_1i_2}^{j_1j_2}(x)-v_{jj_2}^{i_1i_2}(T^\dagger_a)_{j_1}^j\overline{\Phi}^{j_1j_2}_{i_1i_2}(x)\rangle_\epsilon, \nonumber\\
(T_a^\dagger)^{n_2}_{j}\langle \Phi^{n_1j}_{m_1m_2}(y)\rangle_\epsilon&=&-\epsilon\int d^4 x\langle\Phi_{m_1m_2}^{n_1n_2}(y),\overline{v}^{i_1i_2}_{j_1j}(T^\dagger_a)^j_{j_2}\Phi^{j_1j_2}_{i_1i_2}(x)-v_{j_1j_2}^{i_1i}(T_a)_i^{i_2}\overline{\Phi}^{j_1j_2}_{i_1i_2}(x)\rangle_\epsilon,  \nonumber \\
(T_a)_{m_2}^{i}\langle \Phi_{m_1i}^{n_1n_2}(y)\rangle_\epsilon&=&-\epsilon\int d^4 x\langle\Phi_{m_1m_2}^{n_1n_2}(y),\overline{v}^{i_1i}_{j_1j_2}(T_a)^{i_2}_i\Phi_{i_1i_2}^{j_1j_2}(x)-v_{j_1j}^{i_1i_2}(T^\dagger_a)_{j_2}^j\overline{\Phi}^{j_1j_2}_{i_1i_2}(x)\rangle_\epsilon. \nonumber \label{theta24}\\
\end{eqnarray}
As before,  under the action of the diagonal group, the $\epsilon$-term is invariant. So, for the elements of the diagonal group parametrized by $\theta_a^{(1)}=\theta_a^{(\bar{1})}=\theta_a^{(2)}=\theta_a^{(\bar{2})}=\theta_a$,  we take derivatives with respect to  $\theta_a$,  and obtain the extra constraint
\bea\label{diagonalconstraint43}
&&~~~\int d^4x \langle\overline{v}^{i_1i_2}_{jj_2}(T^{\dagger}_a)^{j}_{j_1}\Phi^{j_1j_2}_{i_1i_2}(x)-v^{ii_2}_{j_1j_2}(T_a)^{i_1}_{i}\overline{\Phi}^{j_1j_2}_{i_1i_2}(x)\rangle_{\epsilon}\nonumber\\
&&-\int d^4x \langle\overline{v}^{ii_2}_{j_1j_2}(T_a)^{i_1}_{i}\Phi^{j_1j_2}_{i_1i_2}(x)-v^{i_1i_2}_{jj_2}(T^\dagger_a)^{j}_{j_1}\overline{\Phi}^{j_1j_2}_{i_1i_2}(x)\rangle_{\epsilon}\nonumber\\
&&+\int d^4x \langle\overline{v}^{i_1i_2}_{j_1j}(T^{\dagger}_a)^{j}_{j_2}\Phi^{j_1j_2}_{i_1i_2}(x)-v^{i_1i}_{j_1j_2}(T_a)^{i_2}_{i}\overline{\Phi}^{j_1j_2}_{i_1i_2}(x)\rangle_{\epsilon}\nonumber\\
&&-\int d^4x \langle\overline{v}^{i_1i}_{j_1j_2}(T_a)^{i_2}_{i}\Phi^{j_1j_2}_{i_1i_2}(x)-v^{i_1i_2}_{j_1j}(T^\dagger_a)^{j}_{j_2}\overline{\Phi}^{j_1j_2}_{i_1i_2}(x)\rangle_{\epsilon}=0.
\eea
This constraint  must hold for any configuration of the field $\Phi(y)$, what means that the relation 
\be\label{diagonalconstraint44}
-v_{j_1j_2}^{ii_2}(T_a)_i^{i_1}+v_{jj_2}^{i_1i_2}(T_a)_{j_1}^j-v_{j_1j_2}^{i_1i}(T_a)_i^{i_2}+v_{j_1j}^{i_1i_2}(T_a)_{j_2}^j
=0,
\ee
must be satisfied for a suitable choice of $v$ which breaks the symmetry into the diagonal group. In terms of the Goldstone bosons \eqref{Bdefinition}, the constraint \eqref{diagonalconstraint43} and/or \eqref{diagonalconstraint44} follows from the cancellation \eqref{basicidentitiesgeneral22G.BB2}, which for this case reads
\be\label{Bes}
B_a^1(x)-B_a^{\bar{1}}(x)+B_a^2(x)-B_a^{\bar{2}}(x)=0.
\ee
The only  option for  $v$ to fulfill the constraint \eqref{diagonalconstraint44} is\footnote{See Fig.\ref{simple1}.}
\be\label{symmetrypattern4}
v^{i_1i_2}_{j_1j_2}=v_1\delta^{i_1i_2}_{j_1j_2}+v_2\delta^{i_1i_2}_{j_2j_1}.
\ee
With $v$ as in \eqref{symmetrypattern4}, the Goldstone bosons are 
\bea
B_a^1(x)&=&\big[{\color{red}v_1\big(\Phi_{i_1i}^{j_1i}(x)-\bar{\Phi}_{i_1i}^{j_1i}(x)\big)}+{\color{magenta}v_2\big(\Phi_{ii_1}^{j_1i}(x)-\bar{\Phi}_{i_1i}^{ij_1}(x)\big)\big](T_a)_{j_1}^{i_1}},\nonumber \\
B_a^{\bar{1}}(x)&=&\big[{\color{red}v_1\big(\Phi_{i_1i}^{j_1i}(x)-\bar{\Phi}_{i_1i}^{j_1i}(x)\big)}+{\color{brown}v_2\big(\Phi_{i_1i}^{i j_1}(x)-\bar{\Phi}_{ii_1}^{j_1i}(x)\big)\big](T_a)_{j_1}^{i_1}},\nonumber\\
B_a^2(x)&=&\big[{\color{blue}v_1\big(\Phi_{ii_2}^{ij_2}(x)-\bar{\Phi}_{ii_2}^{ij_2}(x)\big)}+{\color{brown}v_2\big(\Phi_{i_2i}^{ij_2}(x)-\bar{\Phi}_{ii_2}^{j_2i}(x)\big)\big](T_a)_{j_2}^{i_2}},\nonumber\\
B_a^{\bar{2}}(x)&=&\big[{\color{blue}v_1\big(\Phi_{ii_2}^{ij_2}(x)-\bar{\Phi}_{ii_2}^{ij_2}(x)\big)}+{\color{magenta}v_2\big(\Phi_{ii_2}^{j_2i}(x)-\bar{\Phi}_{i_2i}^{ij_2}(x)\big)\big](T_a)_{j_2}^{i_2}},
\eea
where as before we have depicted the terms which cancel in \eqref{Bes} in colors. Notice that {\it all} the Goldstone bosons are needed for the cancellation, and no partial cancellations occur. Equivalently, there is one loop when joining equal colors in the equations. So, there are three independent Goldstone boson matrix fields, matching the number of broken symmetries in the SSB pattern
\be
 U_1(N)\times U_2(N)\times U_{\bar{1}}(N)\times U_{\bar{2}}(N)\longrightarrow U(N).
 \ee
 
 \section{Conclusion and outlook}\label{conclusions}
 The $\epsilon$-term technique has long  been  proven a powerful tool to tackle SSB, here implemented for the first time in tensor models. This technique leads us to identify the Goldstone bosons as matrix fields, which is one of the central results of the paper. It also enables the discussion of the SSB patterns characterized by the tensor $v$.  In this paper, we focus on SSB patterns leading to diagonal subgroups of $G_{d\bar{d}}(N)$, for which the more general $v$ is a linear combination of Kronecker deltas, as in \eqref{vsigma}. In order to understand the intricate relation between the monomial constituents of $v$ and the SSB patterns, we develop a diagrammatic correspondence. The correspondence provides a visual and straightforward way of interpreting the  SSB on the diagrams.  Unexpectedly, from the diagram inspection, we conclude that any SSB pattern can be induced by only two (complex) parameters. 
 
This work comprises a kinematic  study of vacua in tensor models. We mainly focus on the cases where the effective theory is a matrix theory transforming in the adjoint of $U(N)$, although the treatment holds for other scenarios, see appendix \ref{tetraap}. We claim that any tensor theory with SSB always leads to matrix theories. In fact, equation \eqref{genmatrix} is a general result, which holds for any symmetry group and any SSB pattern. There are other attempts in the literature to relate matrix and tensor theories \cite{Diaz:2018zbg, Bonzom:2013lda, Ferrari:2017ryl,Azeyanagi:2017drg,Ferrari:2017jgw, Krishnan:2016bvg, Krishnan:2017ztz,Itoyama:2017xid}, but those are referred to specific models. In contrast, our approach is applicable to any tensor theory as long as the theory presents SSB, which requires that the potential has nonzero stationary points, as in \eqref{solutionN3}.\\

One of the undoubtable succeeds of tensor theories is the simple large $N$ structure of some models, which makes them exactly solvable. However, this might not be the case for all physically relevant models. SSB  provides valuable information of the system, particularly useful when the large $N$ solvability is lost. Additionally,  as SSB generically leads to matrix models, this mechanism could  conceptually clarify the relation between tensor theories and quantum gravity, holography, etc.

Let us now describe a possible scenario. Suppose we have  a tensor theory with a nontrivial potential which presents several stationary points at which the SSB patterns lead to diagonal subgroups of $G_{d\bar{d}}(N)$, hence to $U(N)$ matrix theories. This is shown in Fig.\ref{landscape}, where we have marked the minima.
\begin{figure}[H]
\centering
  \includegraphics[width=.4\textwidth]{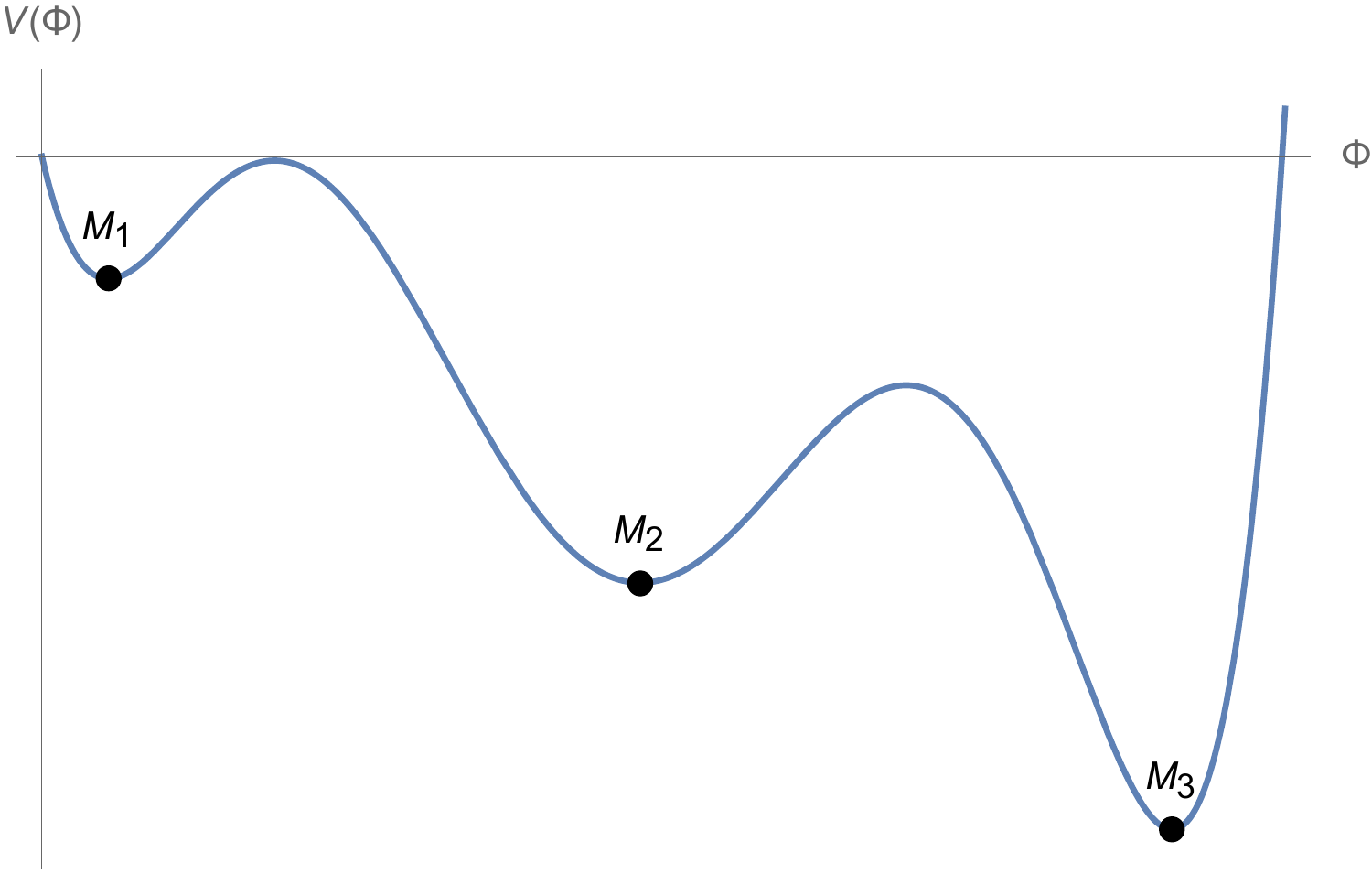}
  \caption{\sl Schematic landscape of matrix theories, with three minima.} \label{landscape}
\end{figure}
Given a theory with a potential as in Fig.\ref{landscape}, by means of SSB we could arrive at any minimum point of the landscape, where different (multi-)matrix theories sit. Notice that the collection of matrix theories is encoded in  an underlying single tensor theory. We believe that this remarkable fact deserves to be explored in depth. On this line, finding specific models which realized  the aforementioned correspondence  is our next goal. As a remark, if the tensor theory were not  solvable in the large $N$, then it could be used to perform non-trivial calculations in the (non-solvable) matrix setups.  

An important point that must be addressed is the dimensionality of spacetime, here taken as $d$. Strictly speaking, spontaneous symmetry breaking happens in QFT and not in quantum mechanics, since the number of degrees of freedom of the latter is finite. Therefore, our results are valid for $d>0+1$. However, we believe that the derivation treated in this paper could be adapted to the cases $d=0+1$ and $d=0$. In those cases, even if we arrived at similar results, it is not obvious to us the physical meaning of the ``matrix Goldstone bosons'' \ref{mainresult}, supposedly massless, since  there is no notion of mass for $d=0+1$ and $d=0$\footnote{For a meaty discussion on the topic see, for instance, \cite{QFTbook}.}. Even in the absence of mass, there is a way of constructing a notion of scale using the parameter $N$, as it has been implemented for the study of the renormalization flow in tensor models with $d=0$  \cite{Eichhorn:2018ylk, Eichhorn:2018phj}. It would be interesting to see if we could apply the same philosophy of those references in order to extend the notion of spontaneous symmetry breaking in tensor theories that we have presented here to quantum mechanics and  random tensor models with $d=0$.

Tensor models in $d=0$ have gained interest lately for describing discretized gravity and providing at the same time the possibility of constructing theories where spacetime is emergent. Our approach in this paper has been orthogonal to that idea. We have focused on connecting tensor theories with string/M-theory. In these theories matrix fields ($d>0+1$) naturally appear and they are usually treated in the large N regime. By applying spontaneous symmetry breaking, matrix fields can be interpreted as coming from a father theory of tensors, which could in turn be thought of as a generalization of string/M-theory.  For instance, ``spontaneous symmetry breaking'' in tensor theories for $d=0+1$ would put in contact our formalism with the realization of M-theory in the formulation \cite{Banks:1996vh}.  This is currently under study.

\section*{Acknowledgements}
We are grateful to Junchen Rong and Robert de Mello Koch for profitable discussions.

\appendix

\section{SSB for the model with tetrahedral vertex}\label{tetraap}
In this appendix we show an example of symmetry breaking occurring in a well-studied tensor model. The pattern of symmetry breaking for this model is different from the patterns
studied in this paper. However, it is worth mentioning it for its potential relevance in future works. It also supports our believe that SSB  is a common feature in tensor models and deserves further
attention.

The real $O_1(N)\times O_2(N)\times O_3(N)$  scalar model with tetrahedral vertex is a $\varphi^4$ real massless scalar tensor model with potential
\be\label{tetra}
V(\varphi)=\varphi_{ijk}\varphi_{ilm}\varphi_{njm}\varphi_{nlk},
\ee
 where the tensor field transforms under $O_1(N)\times O_2(N)\times O_3(N)$ as
\be
\varphi'_{ijk}=(g_1)_{il}(g_2)_{jm}(g_3)_{kn}\varphi_{lmn},\quad g_i\in O_i(N).
\ee
To study SSB we proceed as usual by searching for stationary points of the Lagrangian which happens at the solutions  
\be
\frac{\partial}{\partial\varphi}V(\varphi)\bigg|_{\varphi=v}=0.
\ee
If there are non-trivial solutions of this equation, namely $v\neq 0$, then the theory allows for SSB around the configurations $\varphi(x)$ which
have a non-zero value at infinity\footnote{It was notice in \cite{Klebanov:2016xxf} that the potential \eqref{tetra} has a negative direction. Thus, solution \eqref{solutionN3} is not a minimum. However, as stated in \cite{Weinberg:1996kr, Polchinski},   the discussion of SSB also applies to stationary points of the potential.}.
It is interesting to see that although the usual massless $\phi^4$ theory do not present any SSB the tensor theory does. To see this consider first the case $N=3$. It is not hard to check that a solution of the equation
 \bea \label{solutionN3}
-2v&=&v_{123}=v_{213}=v_{321}=v_{132}=v_{231}=v_{312},\nonumber \\
-2v&=&v_{111}=v_{222}=v_{333}, \nonumber \\
v&=&v_{112}=v_{121}=v_{211}=v_{113}=v_{131}=v_{311}=v_{221}=v_{212}=v_{122}\nonumber \\
&=&v_{223}=v_{232}=v_{322}=v_{331}=v_{313}=v_{133}=v_{332}=v_{323}=v_{233},
\eea
for any $v\in \mathbb{R}$ is a solution of the collection of equations
\be
\frac{\partial}{\partial \varphi_{ijk}}V(v)=v_{imn}v_{ljn}v_{lmk}=0,\qquad i,j,k=1,2,3.
\ee
We would like to emphasize that the existence of a solution like \eqref{solutionN3} is a purely tensor effect, since the only solution of $V'(\phi)=0$ for the scalar case
is $\phi=0$.

Notice that the solution \eqref{solutionN3} is invariant under the action of the diagonal group $S_3$. The diagonal action of $S_N$ on $\varphi$ is defined as
\be\label{phitrans}
\varphi'_{ijk}= \varphi_{\sigma(i)\sigma(j)\sigma(k)},\qquad \sigma\in S_N.
\ee
The invariance of $v$ under $S_3$ is stated as
\be\label{vinv}
v_{ijk}= v_{\sigma(i)\sigma(j)\sigma(k)},\qquad \sigma\in S_3.
\ee
We argue that for configurations $\varphi(x)$ which take values \eqref{solutionN3} at infinity with $v\neq 0$, the theory present SSB and the 
remaining symmetry is precisely $\text{Diag}[S_3]$.

The solution \eqref{solutionN3} can be extended to larger values of $N$ with a simple prescription: for any $O(N)$ we pick a subspace $O(3)$, labeled by 3 different numbers from 1 to $N$.  The solution
 \eqref{solutionN3} applies on that subspace, and $v_{ijk}=0$ for any value of $\{i,j,k\}$ out of the chosen triplet. Moreover, for $N= 6$ we could pick two $O(3)$ subspaces and apply \eqref{solutionN3} with two independent and nonzero values of $v$ on each subspace. In general,  for arbitrary $N$, we could chose $k$ $O(3)$ subspaces, where the solution pattern  \eqref{solutionN3}  applies, and break the symmetry
\begin{multline}\label{SSBforN}
 O_1(N)\times O_2(N)\times O_3(N)\longrightarrow  \\ 
O_1(N-3k)\times O_2(N-3k)\times O_3(N-3k)
\times \text{Diag}[S_3]_1\times \cdots \times \text{Diag}[S_3]_k.
\end{multline}
In order to complete the discussion let us calculate the Goldstone bosons. Without loss of generality we will take $N=3$.  Using the definition \eqref{Bdefinition} adapted to $O(3)$ we may write 
\bea\label{GBO3}
B_a^{(1)}(x)&=&v_{i_1i_2i_3}(T_a)_{i_1k}\varphi_{ki_2i_3}(x)\nonumber \\
B_a^{(2)}(x)&=&v_{i_1i_2i_3}(T_a)_{i_2k}\varphi_{i_1ki_3}(x)\nonumber \\
B_a^{(3)}(x)&=&v_{i_1i_2i_3}(T_a)_{i_3k}\varphi_{i_1i_2k}(x),
\eea
where $T_a$ are the generators of the algebra of $O(3)$. They can be written as
\be
(T_{mn})_{ik}=-\text{i}(\delta_{mk}\delta_{in}-\delta_{nk}\delta_{im}),
\ee
where the index $a=1,2,3$ has been mapped to the  pair $(mn)=(12),(13),(23)$.

As the in the unitary case, the Goldstone bosons in \eqref{GBO3} get arranged into matrices. Using \eqref{genmatrix} adapted to the orthogonal group we may write
\bea\label{GBO31}
(B^{(1)})_{ij}(x)&=&v_{i_1i_2i_3}\big[(T_a)_{i_1k}(T_a)_{ij}\big]\varphi_{ki_2i_3}(x)\nonumber \\
(B^{(2)})_{ij}(x)&=&v_{i_1i_2i_3}\big[(T_a)_{i_2k}(T_a)_{ij}\big]\varphi_{i_1ki_3}(x)\nonumber \\
(B^{(3)})_{ij}(x)&=&v_{i_1i_2i_3}\big[(T_a)_{i_3k}(T_a)_{ij}\big]\varphi_{i_1i_2k}(x).
\eea
The multiplication rule for the orthogonal group is
\be\label{ruleortho}
(T_a)_{i_1k}(T_a)_{ij}=(T_{mn})_{i_1k}(T_{mn})_{ij}=-(\delta_{kj}\delta_{i_1i}-\delta_{ji_1}\delta_{ki}),
\ee
which implemented on \eqref{GBO31} yields
\bea\label{GBO32}
(B^{(1)})_{ij}(x)&=&-(v_{ii_2i_3}\varphi_{ji_2i_3}(x)-v_{ji_2i_3}\varphi_{ii_2i_3}(x))\nonumber \\
(B^{(2)})_{ij}(x)&=&-(v_{i_1ii_3}\varphi_{i_1ji_3}(x)-v_{i_1ji_3}\varphi_{i_1ii_3}(x))\nonumber \\
(B^{(3)})_{ij}(x)&=&-(v_{i_1i_2i}\varphi_{i_1i_2j}(x)-v_{i_1i_2j}\varphi_{i_1i_2i}(x)).
\eea
Since we are considering $O(3)$ the continuous symmetry gets completely broken, leading to the three independent
Goldstone boson matrices \eqref{GBO32}. Notice that the nine generators of the original symmetry group $O(3)\times O(3)\times O(3)$ match the number of independent components of the three $3\times 3$-antisymmetric matrices in \eqref{GBO32}.

Although there is no continuous symmetry left, it still  remains a discrete $S_3$ symmetry. Interestingly, we can track how the Goldstone bosons transform under this symmetry. Let us see how it goes for $B^{(1)}$. First, realize that  
\begin{multline}\label{btrans}
(B^{(1)})_{ij}(x)=-(v_{ii_2i_3}\varphi_{ji_2i_3}(x)-v_{ji_2i_3}\varphi_{ii_2i_3}(x))\\
=-(v_{i\sigma(i_2)\sigma(i_3)}\varphi_{j\sigma(i_2)\sigma(i_3)}(x)-v_{j\sigma(i_2)\sigma(i_3)}\varphi_{i\sigma(i_2)\sigma(i_3)}(x)).
\end{multline}
The second equality in \eqref{btrans} is just a rearrangement of the sum over $i_2$ and $i_3$ induced by the permutation $\sigma$. According to \eqref{phitrans} the transformation of $B^{(1)}$ under the diagonal $S_3$ is 
\begin{multline}
 (B^{\prime(1)})_{ij}(x)=(B^{(1)})_{\sigma(i)\sigma(j)}(x)=\\
 -(v_{\sigma(i)\sigma(i_2)\sigma(i_3)}\varphi_{\sigma(j)\sigma(i_2)\sigma(i_3)}(x)-v_{\sigma(j)\sigma(i_2)\sigma(i_3)}\varphi_{\sigma(i)\sigma(i_2)\sigma(i_3)}(x)).
 \end{multline}
Using the invariance of $v$  \eqref{vinv}, the transformed Goldstone boson reads
\be 
 (B^{\prime(1)})_{ij}(x)=-(v_{ii_2i_3}\varphi_{\sigma(j)\sigma(i_2)\sigma(i_3)}(x)-v_{ji_2i_3}\varphi_{\sigma(i)\sigma(i_2)\sigma(i_3)}(x)).
 \ee
This transformation is the analog to the adjoint action \eqref{Ztransform} for the unitary group. \\

The procedure described above can be straightforwardly extended to the general SSB \eqref{SSBforN}.  It would be interesting to study the effective theory related to those symmetry breaking patterns. We leave it for a future work.

\end{document}